%% file: EXO-16-018_temp.tex
\pdfoutput=1

\documentclass[11pt,twoside,a4paper,cmspaper,final,collab]{cms-tdr}
\def\svnVersion{365198}\def\cmsCernNoTag{CERN-EP/2016-154}\def\cmsCernDate{\today}\def\cmsMessage{Published in Physical Review Letters as \href{http://dx.doi.org/10.1103/PhysRevLett.117.051802}{\doi{10.1103/PhysRevLett.117.051802.}}}

\begin{document}\cmsNoteHeader{EXO-16-018}

\hyphenation{had-ron-i-za-tion}
\hyphenation{cal-or-i-me-ter}
\hyphenation{de-vices}
\RCS$Revision: 358776 $
\RCS$HeadURL: svn+ssh://svn.cern.ch/reps/tdr2/papers/EXO-16-018/trunk/EXO-16-018.tex $
\RCS$Id: EXO-16-018.tex 358776 2016-07-22 17:38:43Z alverson $
\newlength\cmsFigWidth
\newlength\cmsSmallFigWidth
\newlength\cmsFigMedWidth
\newlength\cmsFigDoubleWidth
\ifthenelse{\boolean{cms@external}}{\setlength\cmsSmallFigWidth{0.85\columnwidth}}{\setlength\cmsSmallFigWidth{0.38\textwidth}}
\ifthenelse{\boolean{cms@external}}{\setlength\cmsFigWidth{0.98\columnwidth}}{\setlength\cmsFigWidth{0.49\textwidth}}
\ifthenelse{\boolean{cms@external}}{\setlength\cmsFigMedWidth{0.98\columnwidth}}{\setlength\cmsFigMedWidth{0.75\textwidth}}
\ifthenelse{\boolean{cms@external}}{\setlength\cmsFigDoubleWidth{0.98\columnwidth}}{\setlength\cmsFigDoubleWidth{0.9\textwidth}}
\ifthenelse{\boolean{cms@external}}{\providecommand{\cmsLeft}{top\xspace}}{\providecommand{\cmsLeft}{left\xspace}}
\ifthenelse{\boolean{cms@external}}{\providecommand{\cmsRight}{bottom\xspace}}{\providecommand{\cmsRight}{right\xspace}}
\ifthenelse{\boolean{cms@external}}{\providecommand{\suppMaterial}{the Supplemental Material~\cite{suppMatRef}}\xspace}{\providecommand{\suppMaterial}{Appendix~\ref{app:supp_material}\xspace}}
\ifthenelse{\boolean{cms@external}}{\providecommand{\CL}{C.L.\space}}{\providecommand{\CL}{CL\xspace}}
\providecommand{\PX}{\ensuremath{\mathrm{X}}\xspace}
\providecommand{\ktild}{\ensuremath{\tilde{k}}\xspace}
\providecommand{\sqrts}{\ensuremath{\sqrt{s}}\xspace}
\providecommand{\mgg}{\ensuremath{m_{\gamma\gamma}}\xspace}
\providecommand{\Zee}{\ensuremath{\cPZ \to \Pep\Pem}\xspace}
\providecommand{\gaga}{\ensuremath{\gamma \gamma}\xspace}
\providecommand{\mX}{\ensuremath{m_{\PX}}\xspace}
\providecommand{\GammaOm}{\ensuremath{\Gamma_{\PX} / m_{\PX}}\xspace}
\providecommand{\Rnine}{\ensuremath{R_{9}}\xspace}
\providecommand{\scEta}{\ensuremath{\abs{\eta_{C}}}\xspace}
\providecommand{\RS}{{\textsc{RS}}\xspace}
\providecommand{\mylumiRone}{19.7\fbinv}
\providecommand{\mylumi}{3.3\fbinv}
\providecommand{\lumiBon}{2.7\fbinv}
\providecommand{\lumiBoff}{0.6\fbinv}
\providecommand{\bon}{3.8\unit{T}}
\providecommand{\boff}{0\unit{T}}
\providecommand{\bonNoUnit}{3.8\xspace}
\providecommand{\boffNoUnit}{0\xspace}
\providecommand{\bUnit}{\unit{T}}
\providecommand{\MGAMC}{\textsc{madgraph5\_amc@nlo}\xspace}

\cmsNoteHeader{EXO-16-018}
\title{Search for resonant production of high-mass photon pairs in proton-proton collisions at
  \texorpdfstring{$\sqrt{s}=8$ and $13\TeV$}{sqrt(s) = 8 and 13 TeV}}

\date{\today}

\abstract{
A search for the resonant production of high-mass photon pairs is presented. The analysis is
based on samples of proton-proton collision data collected by the CMS experiment at
center-of-mass energies of 8 and 13\TeV, corresponding to integrated
luminosities of 19.7 and 3.3\fbinv, respectively.
The interpretation of the search results focuses on spin-0 and spin-2 resonances with masses between 0.5 and
4\TeV and with widths, relative to the mass, between $1.4\times 10^{-4}$ and
$5.6\times10^{-2}$. Limits are set on scalar resonances produced through gluon-gluon fusion, and
on Randall--Sundrum gravitons. A modest excess of events compatible with a narrow resonance
with a mass of about 750\GeV is observed. The local significance of the excess
is approximately 3.4 standard deviations. The significance is reduced to 1.6 standard
deviations once the effect of searching under multiple signal hypotheses is
considered. More data are required to determine  the origin of this excess.
}

\hypersetup{%
pdfauthor={CMS Collaboration},%
pdftitle={Search for resonant production of high-mass photon pairs in proton-proton collisions at
  sqrt(s) = 8 and 13 TeV},%
pdfsubject={CMS},%
pdfkeywords={CMS, LHC, Extra dimensions, Randall-Sundrum, heavy resonance, spin-0,
  proton-proton collision, diphoton}
}

\maketitle

The resonant production of high-mass photon pairs is a prediction that arises in several
extensions of the standard model (SM) of particle physics.
The spin of a resonance decaying to two photons
must be either 0 or an integer greater than or equal to 2~\cite{Landau,PhysRev.77.242}.
Spin-0 resonances decaying to two photons are predicted by models with nonminimal
Higgs sectors~\cite{Branco:2011iw,Lee:1973iz},   while spin-2 resonances decaying to two
photons can arise in models with additional space-like
dimensions~\cite{Randall:1999ee}.

In this Letter, we report on a search for high-mass resonances that decay to photon
pairs. The search is based on proton-proton (pp) collision data collected in 2012 and 2015
by the CMS experiment at the CERN LHC at $\sqrt{s}=8$ and 13\TeV, respectively,
corresponding to integrated luminosities of 19.7 and 3.3\fbinv.
Events with at least two reconstructed photon candidates are selected and a search is
performed in the diphoton mass spectrum for a localized excess of events consistent with
the resonant production of a photon pair.
The results are obtained through a combined analysis of the 8 and 13\TeV data. The data
are interpreted in terms of spin-0 resonances produced through gluon-gluon fusion and
in terms of spin-2 graviton resonances in Randall--Sundrum (\RS)
models~\cite{Davoudiasl:1999jd}. In these models, the spin-2 resonances are produced
through both gluon-gluon fusion and quark annihilation,
with the first mechanism accounting for roughly 90\% of the production cross section.
A portion of the 13\TeV data (0.6\fbinv) was collected when the CMS magnet was
off (\boff), because of an intermittent problem, subsequently rectified, with the cryogenic
system. The remainder of the 13\TeV data, and all of the 8\TeV data, were recorded  with
the magnet at its operational field strength (\bon).

Previous LHC searches for spin-0 resonances decaying to two photons
were performed at $\sqrt{s}=8\TeV$~\cite{ATLAS-dipho-scalar-8TeV,CMS-dipho-8TeV}, and for
spin-2 resonances  decaying to a pair of photons, leptons, jets, or vector bosons at
$\sqrt{s}=7$, 8, and
13\TeV~\cite{ATLAS-dipho-RS-8TeV,ATLAS-dipho-RS-7TeV,ATLAS-dilep-8TeV,ATLAS-dilep-7TeV,ATLAS-dibos-7TeV,ATLAS-dibos-WZ-7TeV,CMS-dipho-8TeV,CMS-dipho-RS-7TeV,CMS-dilep-8TeV,CMS-dilep-7-8TeV,CMS-dijet-13TeV,CMS-dijet-8TeV,CMS-dijet-scout-8TeV,CMS-qbh-8TeV,CMS-dibos-boost-8TeV,CMS-dibos-7TeV,CMS-dibos-ZZ-7TeV}.
The results presented in this Letter exceed the sensitivity of these previous studies, for
spin-0 and spin-2 resonance masses above 500\gev.

A detailed description of the CMS detector, together with a definition of the coordinate
system used and the relevant kinematic variables, can be found
elsewhere~\cite{Chatrchyan:2008zzk}.
The central feature of the CMS apparatus is a superconducting solenoid of 6\unit{m}
internal diameter. Within the solenoid volume are a silicon pixel and strip tracker, a
lead tungstate crystal electromagnetic calorimeter (ECAL), and a brass and scintillator
hadron calorimeter, each composed of a barrel and two endcap sections. Muons are measured
in gas-ionization detectors embedded in the steel flux-return yoke outside the solenoid.
The ECAL consists of about 76\,000 $\rm{PbWO_4}$ crystals that have transverse sizes
approximately matching the Moli\`{e}re radius of the material.
The ECAL barrel (EB), covering the pseudorapidity ($\eta$) region $\abs{\eta}<1.45$, has a
granularity $\Delta\eta\times\Delta\phi=0.0174\times 0.0174$,  with $\phi$ the azimuthal
angle. The ECAL endcaps (EE), which extend the coverage to $\abs{\eta}<3.0$, have a
granularity that increases progressively up to $\Delta\eta\times\Delta\phi=0.05\times
0.05$.
The particle-flow algorithm~\cite{CMS-PAS-PFT-09-001,CMS-PAS-PFT-10-001}
reconstructs and identifies each individual particle with an optimised combination of
information from the various elements of the CMS detector. Particle candidates are
classified as either muons, electrons, photons, $\tau$ leptons, charged hadrons, or
neutral hadrons.

Simulated signal samples of spin-0 and spin-2 resonances decaying to two photons
are generated at leading order (LO) with the $\PYTHIA\,8.2$~\cite{Sjostrand:2014zea} event
generator, using the NNPDF2.3~\cite{Ball:2012cx} parton distribution functions (PDFs),
with values of the resonance mass \mX in the range $0.5<\mX<4\TeV$
and for three values of the relative width \GammaOm:
$1.4\times10^{-4},1.4\times10^{-2}$ and $5.6\times10^{-2}$.
For the RS graviton model, where $\GammaOm=1.4\,\ktild^2$~\cite{Davoudiasl:1999jd},
this corresponds to dimensionless coupling values $\ktild=0.01,0.1$, and $0.2$.
The chosen relative widths correspond, respectively, to resonances much narrower than,
comparable to, and significantly wider than the detector resolution.
The principal SM background processes, namely the direct production of two photons
(\gaga), the production
of $\gamma+$jets events in which jet fragments are misidentified  as photons, and
the production of multijet events with misidentified jet fragments, are simulated
with the $\SHERPA\,2.1$~\cite{Gleisberg:2008ta}, $\MGAMC\,2.2$~\cite{Alwall:2014hca}
(interfaced with $\PYTHIA\,8.2$ for parton showering and hadronization), and
$\PYTHIA\,8.2$ generators, respectively.
For all simulated samples, the detector response
is modeled with the $\GEANT4$ package~\cite{Agostinelli:2002hh}.
The kinematic requirements and the identification criteria described below are determined
using the simulated signal and background samples and are finalized prior to inspecting the
diphoton mass data distribution in the search region.

For the 8\TeV data, the results of Ref.~\cite{CMS-dipho-8TeV} are used in the present
study to place limits on resonances
with $\mX \leq 850\GeV$.  In this paper, we extend these 8\TeV limits to masses $\mX > 850\GeV$
using an analysis similar to the 13\TeV one.
In the following, we first describe the 13\TeV analysis, then the manner in which the
8\TeV analysis differs.

For the $\mathrm{B}=\bonNoUnit$ (\boffNoUnit)\bUnit data at 13\TeV, the trigger selection
requires at least two
photon  candidates, each with transverse momentum \pt above 60 (40)\GeV. For each photon
candidate, the ratio of the energy deposited in the hadron calorimeter
to the photon energy (H/E ratio) is required to be less than 0.15. For
resonances with $\mX>0.5\TeV$, the trigger selection is fully efficient.

In the subsequent analysis, photons are reconstructed by clustering
spatially correlated energy deposits in the ECAL. To obtain the best energy resolution,
the ECAL signals are calibrated and
corrected for the variation of the crystal transparency during the
data collection period~\cite{CMS:EGM-14-001}. The energies of the photon candidates are
estimated with a multivariate regression technique~\cite{CMS:EGM-14-001}.
For the \bon data, the interaction vertex, i.e., the pp collision point from
which the photons are assumed to originate, is selected using the
algorithm described in Ref.~\cite{Khachatryan:2014ira}. For resonances with
$\mX>500\GeV$, the  fraction of events in which the interaction vertex is correctly
assigned is estimated from simulation to be approximately 90\%.
For the \boff data, the interaction vertex is identified as the reconstructed vertex
with the largest number of charged
tracks, yielding an estimated probability for the correct assignment of about 60\%. The
direction of a photon candidate's momentum is computed taking as the origin the position of
the chosen interaction vertex.
Corrections to account for residual differences in the photon energy scale and resolution
between the data and simulation are
determined using \Zee events, through the procedure described in
Ref.~\cite{CMS:EGM-14-001}. For the \bonNoUnit (\boffNoUnit)\bUnit data, energy scale and resolution
corrections are derived in eight (four) bins defined in terms of the \Rnine variable,
which is the ratio of the energy deposited in the central $3{\times}3$ crystal matrix to
the full cluster energy, and of the \scEta variable, which is the absolute value of the
pseudorapidity
of the cluster with respect to the center of the detector. The energy scale correction
factors measured for the \bon data are found to be about 1\% higher than the \boff factors, while
similar values are measured for the resolution corrections.
The variation of the corrections in the EB (EE) region is assessed as a function of \pt up
to $\pt\approx150$ (100)\GeV using \Zee data, and is found to be 0.5
(0.7)\% or less for both the 3.8 and 0\unit{T} data.

Photon candidates are subject to additional identification requirements. The H/E
ratio of the candidates must lie below 0.05. For the \bonNoUnit (\boffNoUnit)\bUnit data,
the size of the
electromagnetic clusters in $\eta$ ($\eta$ and $\phi$)~\cite{CMS:EGM-14-001}
is required to be compatible with that expected for a prompt photon, i.e., a photon
produced directly in a hard-scattering process.
For candidates in the \bon sample, the scalar \pt sum of additional photons in a cone of radius
$R=\sqrt{\smash[b]{(\Delta\eta)^2+(\Delta\phi)^2}}=0.3$ around the photon direction, corrected to
account for the contributions from  extraneous pp collisions in the same or nearby proton
bunch crossing, must be less than 2.5\GeV.
For the \boff sample, the analogous sum must be less than 3.6\,(3.0)\GeV for the EB (EE) candidates.
For the \bon data, we additionally require the scalar \pt sum of the charged hadrons within a
cone of radius $R=0.3$ around the photon direction to be less than 5\GeV, and for the
\boff data the number of charged hadrons within this cone, excluding an inner cone of
radius $R=0.05$,  to be 3 or less.
The photon isolation requirement for the \boff data is less stringent than that for the \bon
data to compensate for the additional selection criterion for the \boff data based on the size of
the shower profile in the azimuthal direction.
Photon candidates associated with an electron track that itself is not consistent with a
photon conversion are rejected.

For the \bon data, the efficiency of the identification criteria for prompt isolated
photon candidates in the barrel (endcaps) is above 90 (85)\% for the kinematic range
considered in the analysis.
For the \boff data, the corresponding efficiency exceeds 85 (70)\%.
The identification and trigger efficiencies are measured, as a function of \pt,  using
data events containing a Z boson decaying to a pair of electrons, or to a pair of electrons
or muons in association with a photon~\cite{CMS:EGM-14-001}. The efficiencies from data
are found to be consistent with those from simulation.

In each event, photon candidates with $\pt>75\GeV$ are grouped in all possible pairs.
We require $\scEta<2.5$ for each candidate in the pair and $\scEta<1.44$ for at least one
of them. Candidates in the region $1.44 < \scEta < 1.57$ are rejected because of
difficulties in modeling the photon reconstruction efficiency in the transition region
between the barrel and endcap detectors.
The invariant mass \mgg of the pair is required to exceed 230\GeV. For
events in which one photon candidate is reconstructed in an endcap, \mgg must exceed 330\GeV. The
fraction of events in which more than one photon pair satisfies all the selection
criteria is roughly 1\%. In these cases, only the pair with the largest
photon scalar \pt sum is retained.

Photon pairs are divided into two categories, denoted by ``EBEB'' when both photons are
reconstructed in the ECAL barrel and by ``EBEE'' when one of the two photons is reconstructed in
an ECAL endcap. Each category is further divided into events recorded at 3.8 and
0\unit{T}.

For the \bonNoUnit (\boffNoUnit)\bUnit analysis, the overall signal selection efficiency
varies between 0.5--0.7
(0.4--0.5), depending on the signal hypothesis.
Because of the different angular distribution of the decay products,
the kinematic
acceptance for the \RS graviton resonances is lower than for scalar resonances; for
$\mX<1\TeV$ the reduction is approximately 20\%. The two acceptances become similar for
$\mX > 3\TeV$.
About 90 (80)\% of the background events in the EBEB (EBEE) sample arises from the \gaga
process. These results, estimated from simulation, are validated for the \bon analysis
using the method described in Ref.~\cite{smdipho_7TeV}.

The principal difference between the 8\TeV analysis described in Ref.~\cite{CMS-dipho-8TeV}
(used here in the search for resonances with $\mX\leq 850\GeV$) and the 13\TeV
analysis described above is that, in the former, the events are further categorized according to
the \Rnine value of the photon candidates. Specifically, events are categorized as having
either $\min(\Rnine)>0.94$ or $\mathrm{min}(\Rnine)\leq0.94$, where
$\min(\Rnine)$ is the smaller of the
two \Rnine values in the photon pair. To search for resonances with $\mX>850\GeV$ in the
8\TeV data, we select photons with $\pt>80\GeV$ that satisfy the ``loose'' identification
criteria of Ref.~\cite{CMS:EGM-14-001} and require that there be an EBEB photon pair with
$\mgg>300\GeV$. We do not include EBEE photon pairs in this case for reasons of
simplicity, because such events would have improved the analysis sensitivity by at most a
few percent.

\begin{figure*}[hbt]
    \centering
    \includegraphics[width=\cmsSmallFigWidth]{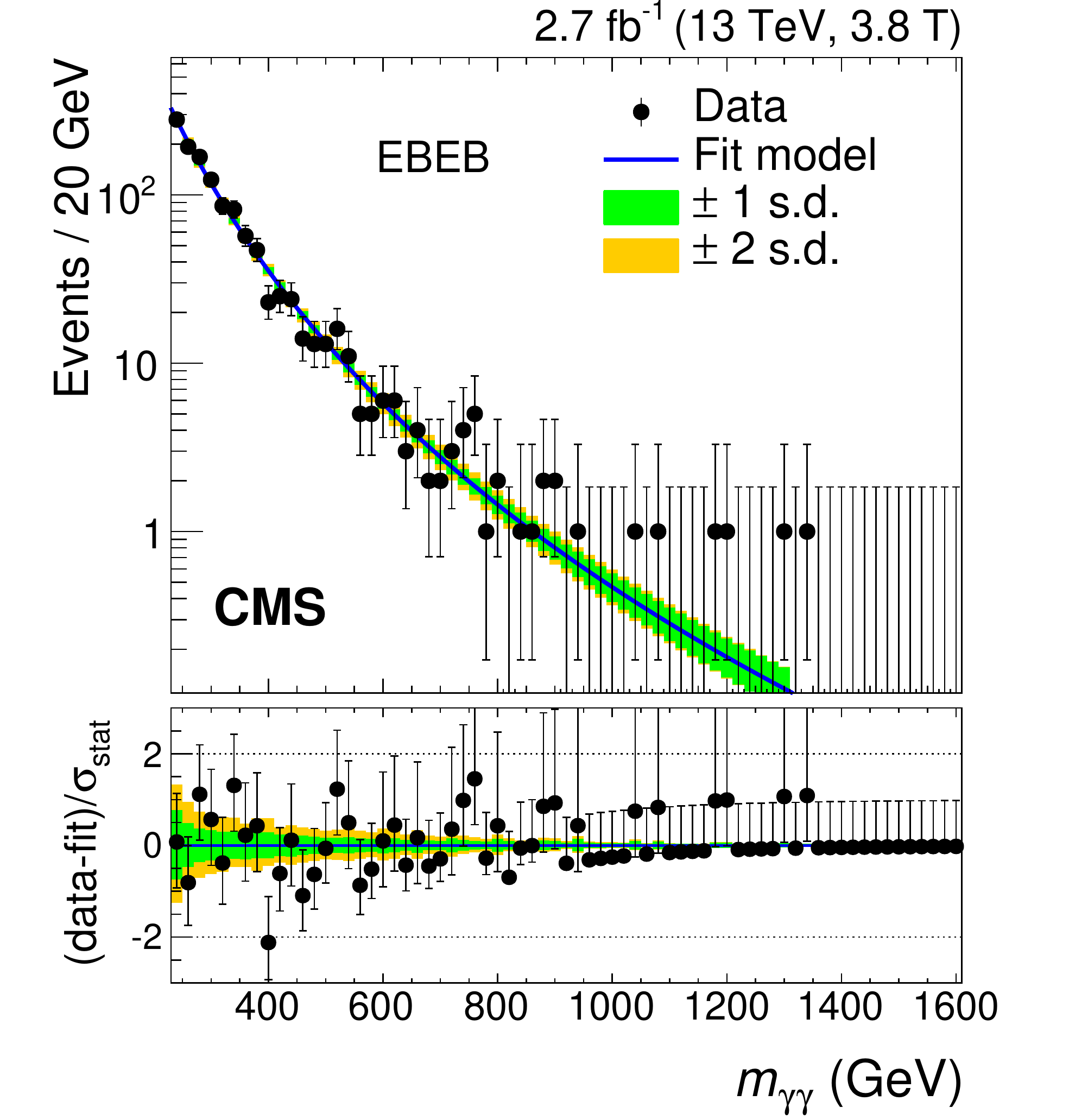}
    \includegraphics[width=\cmsSmallFigWidth]{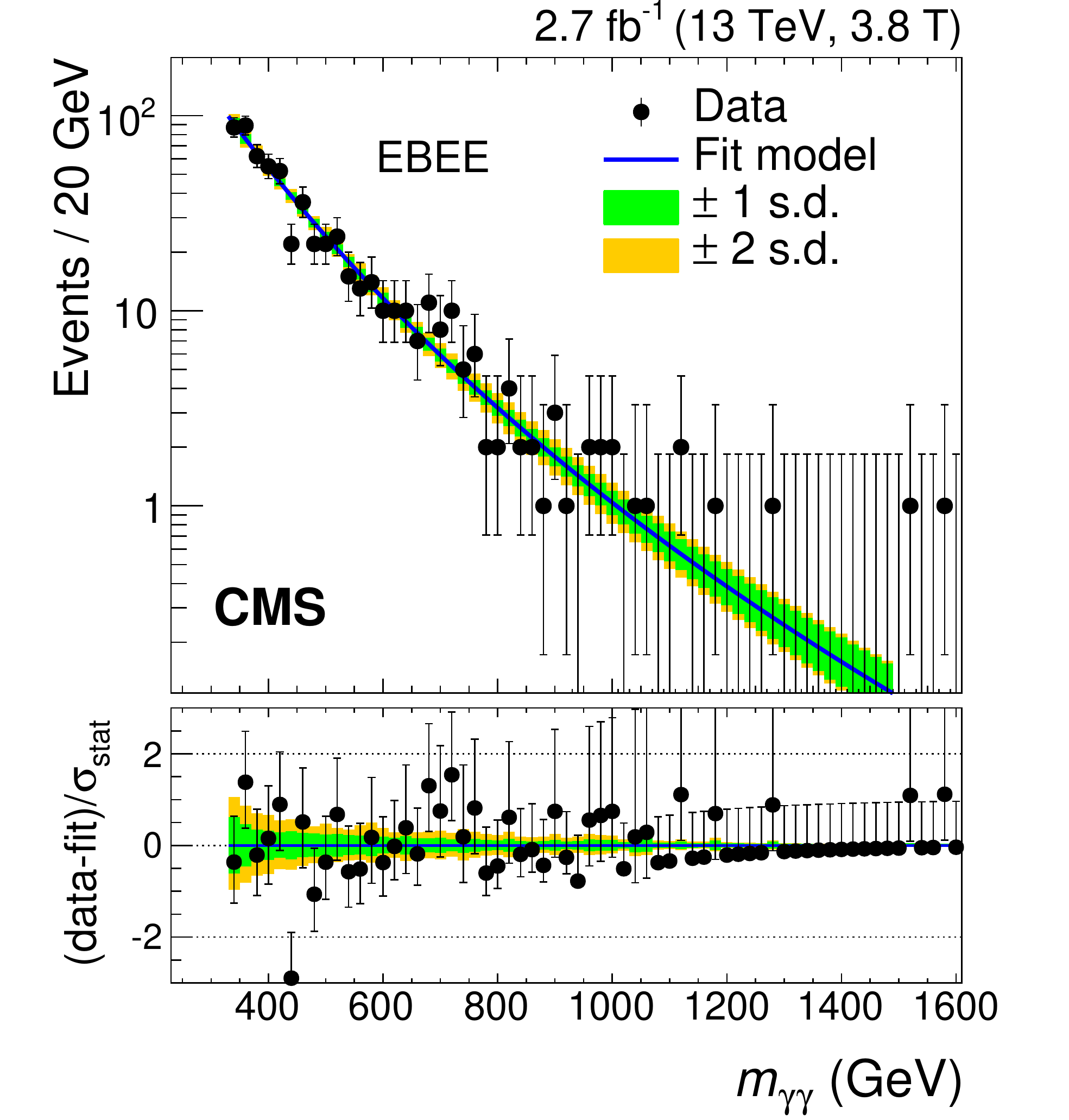} \\
    \includegraphics[width=\cmsSmallFigWidth]{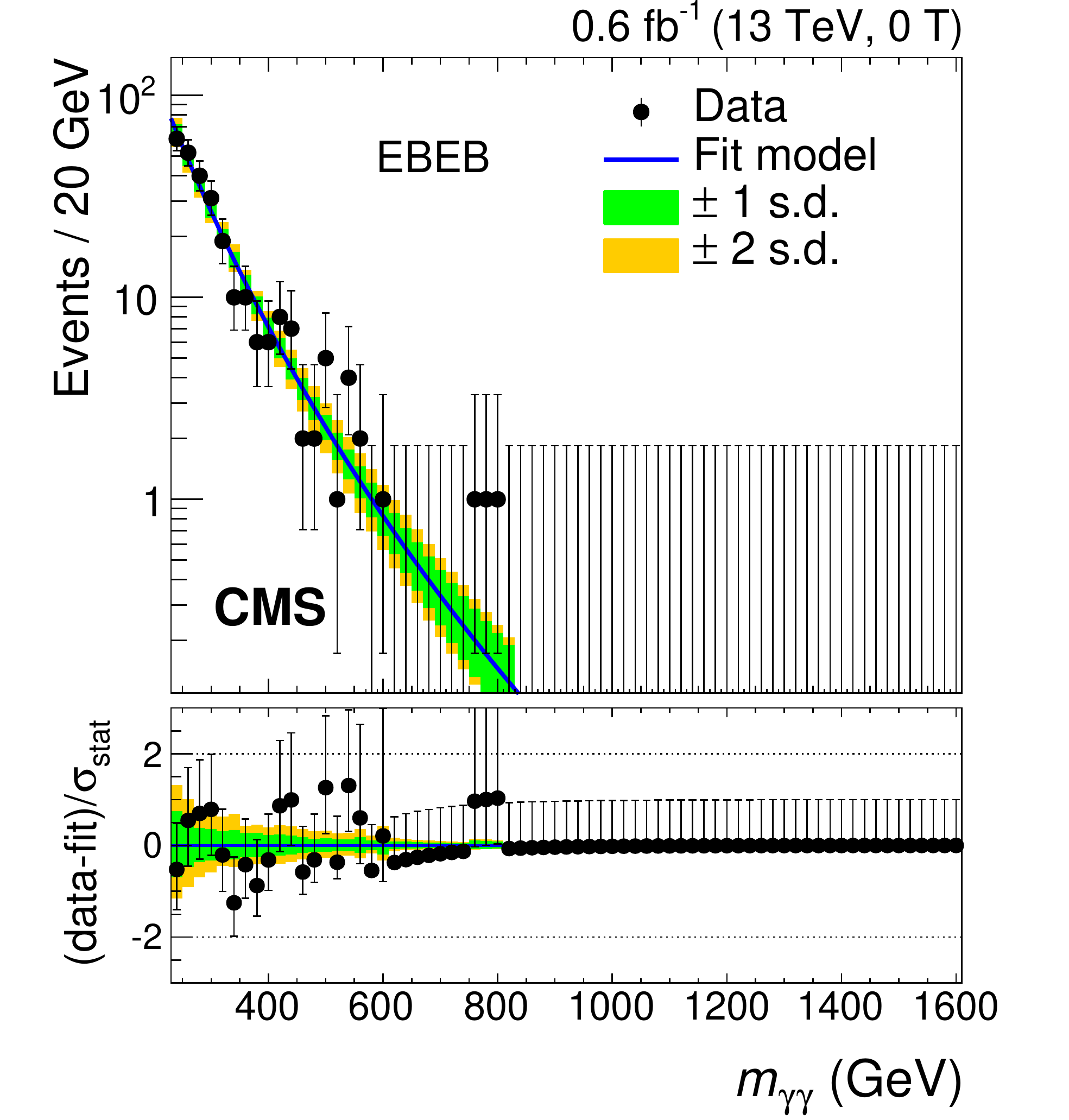}
    \includegraphics[width=\cmsSmallFigWidth]{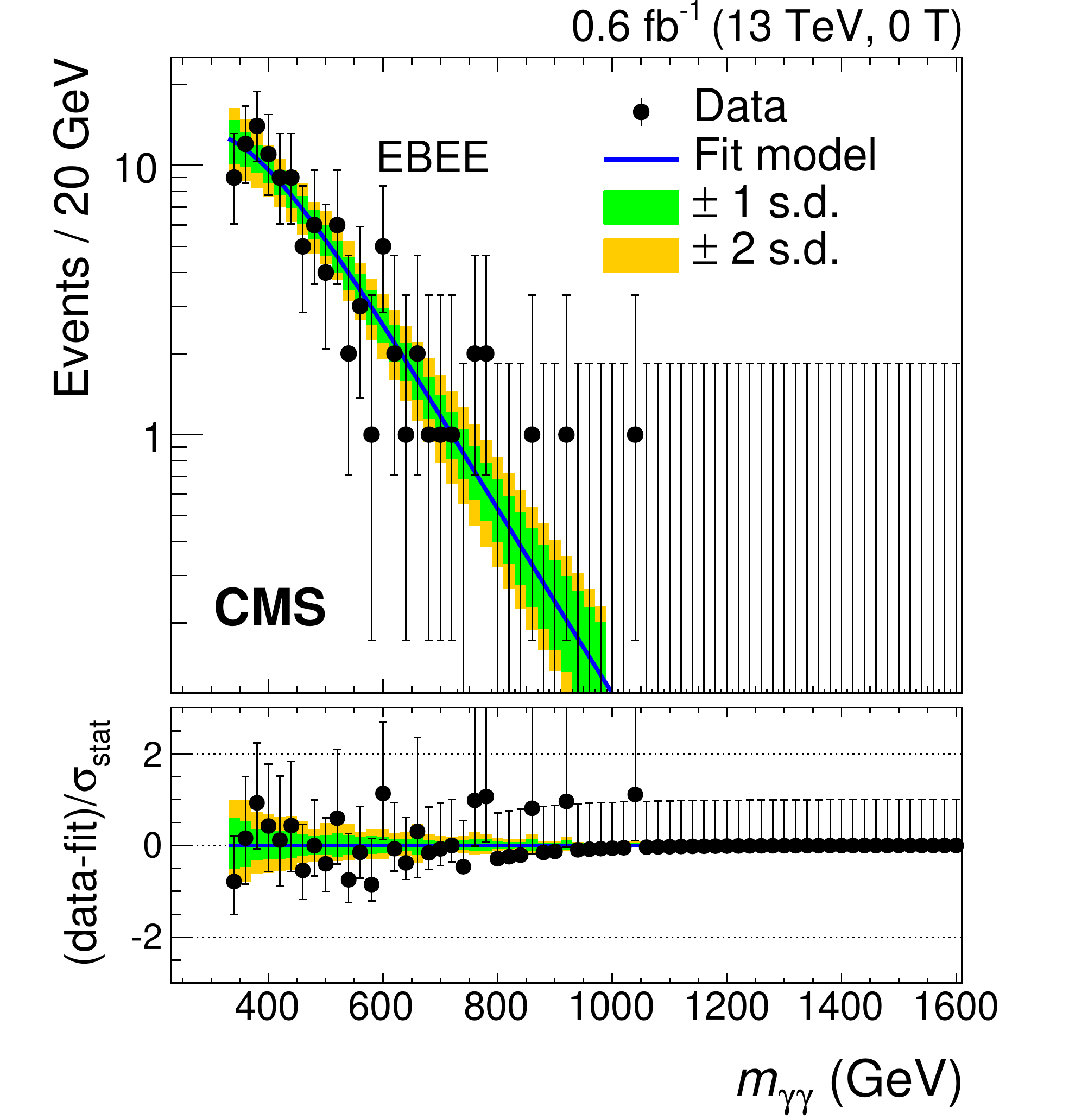}
\caption{
  Observed diphoton invariant mass \mgg spectra for the event categories
  used in the analysis of the 13\TeV data:
  (upper row) magnetic field strength $\mathrm{B}=\bon$;
  (lower row) $\mathrm{B}=\boff$; (left column) both photons in the ECAL barrel
  detector, (right column) one photon in the ECAL barrel detector and the other
  in an ECAL endcap detector. No event with $\mgg>1600\GeV$ is selected in the analysis.
  The results of a likelihood fit to
  the background-only hypothesis are also shown.
  The shaded regions show the 1 and 2 standard deviation uncertainty bands associated with
  the fit, and reflect the statistical uncertainty of the data.
  The lower panels show the difference between the data and fit, divided by the
  statistical uncertainty in the data points.
  \label{fig:fits_13TeV}
}
\end{figure*}

\begin{figure*}[hbt]
    \centering
    \includegraphics[width=\cmsSmallFigWidth]{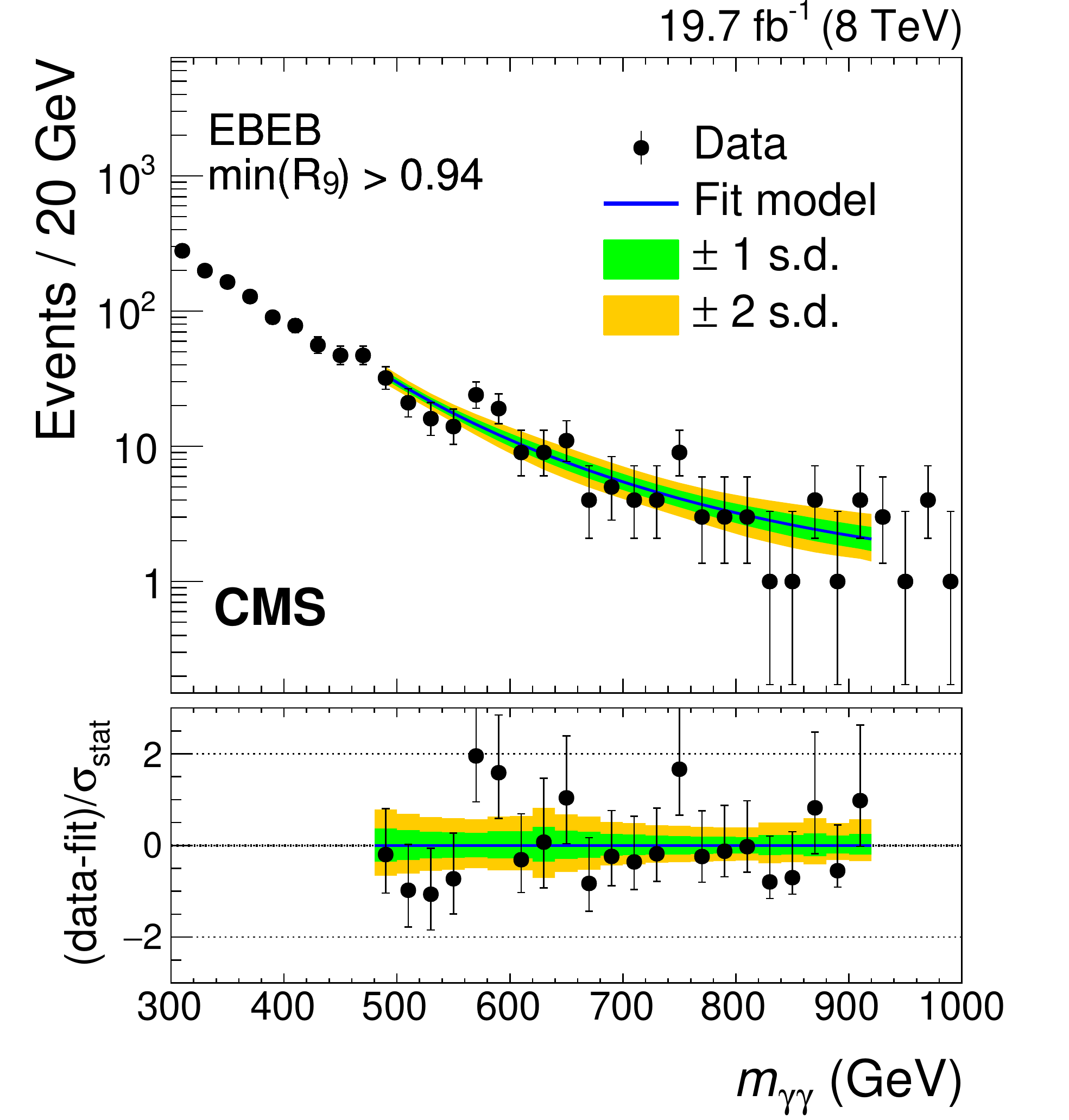}
    \includegraphics[width=\cmsSmallFigWidth]{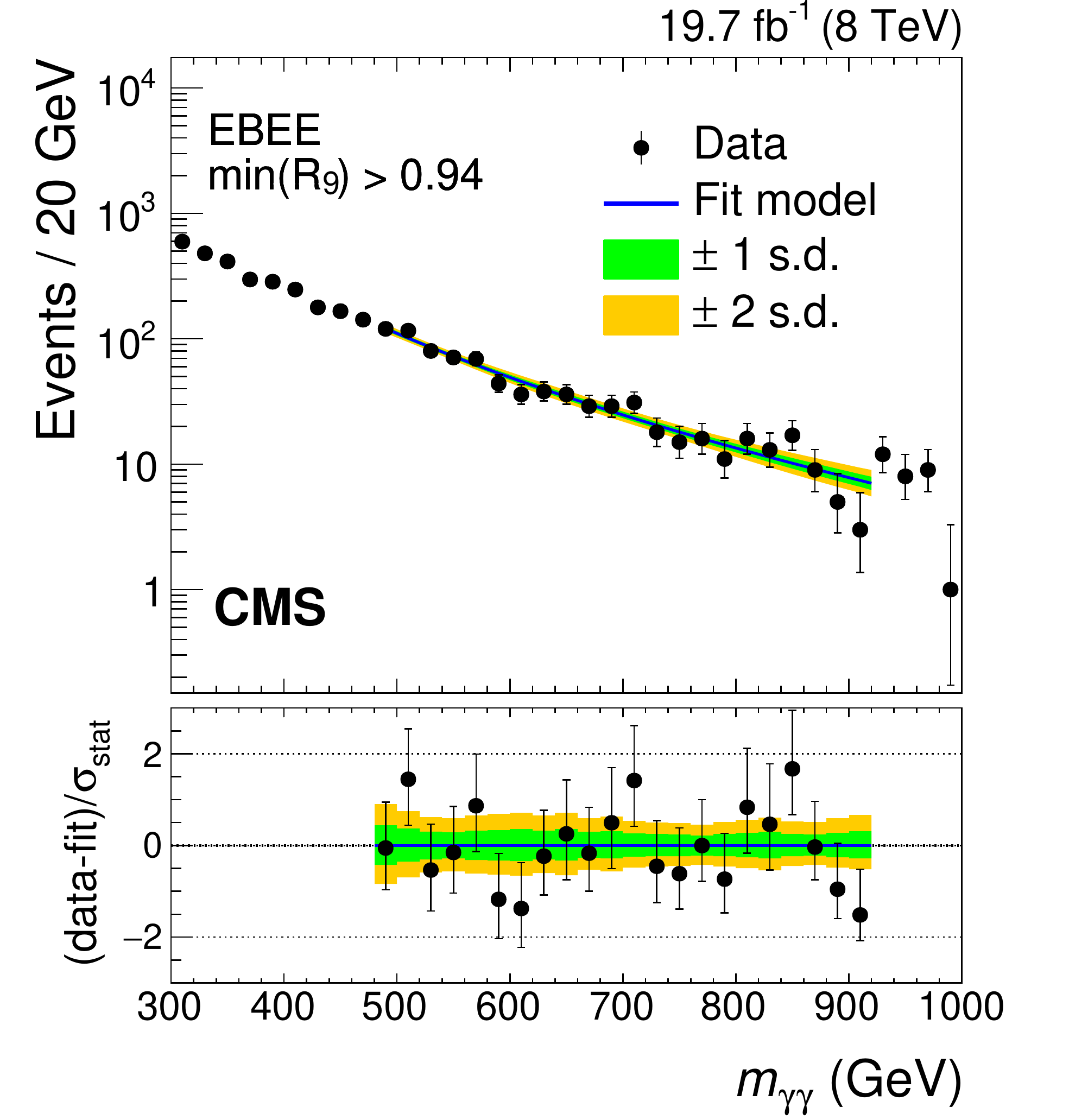} \\
    \includegraphics[width=\cmsSmallFigWidth]{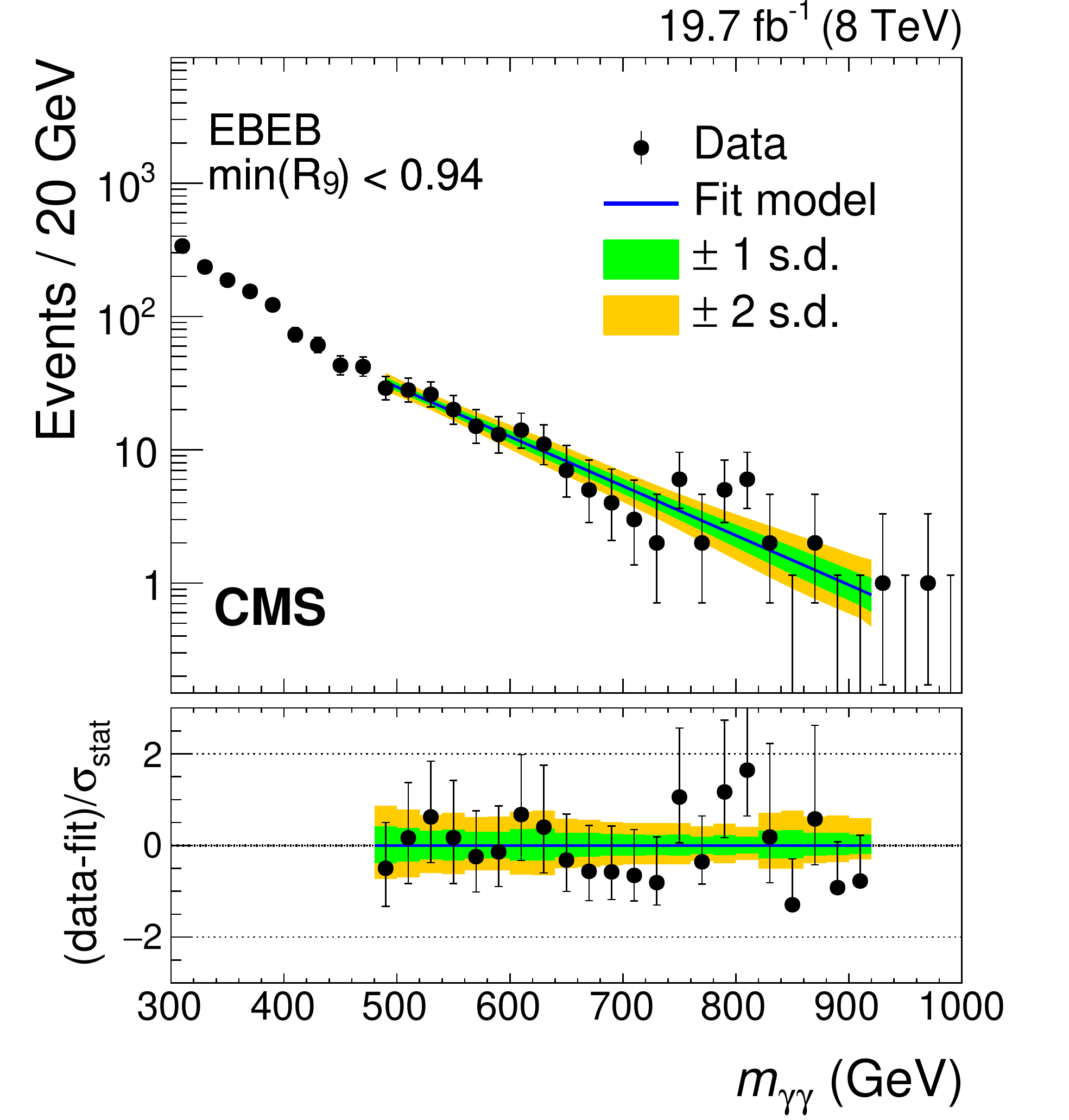}
    \includegraphics[width=\cmsSmallFigWidth]{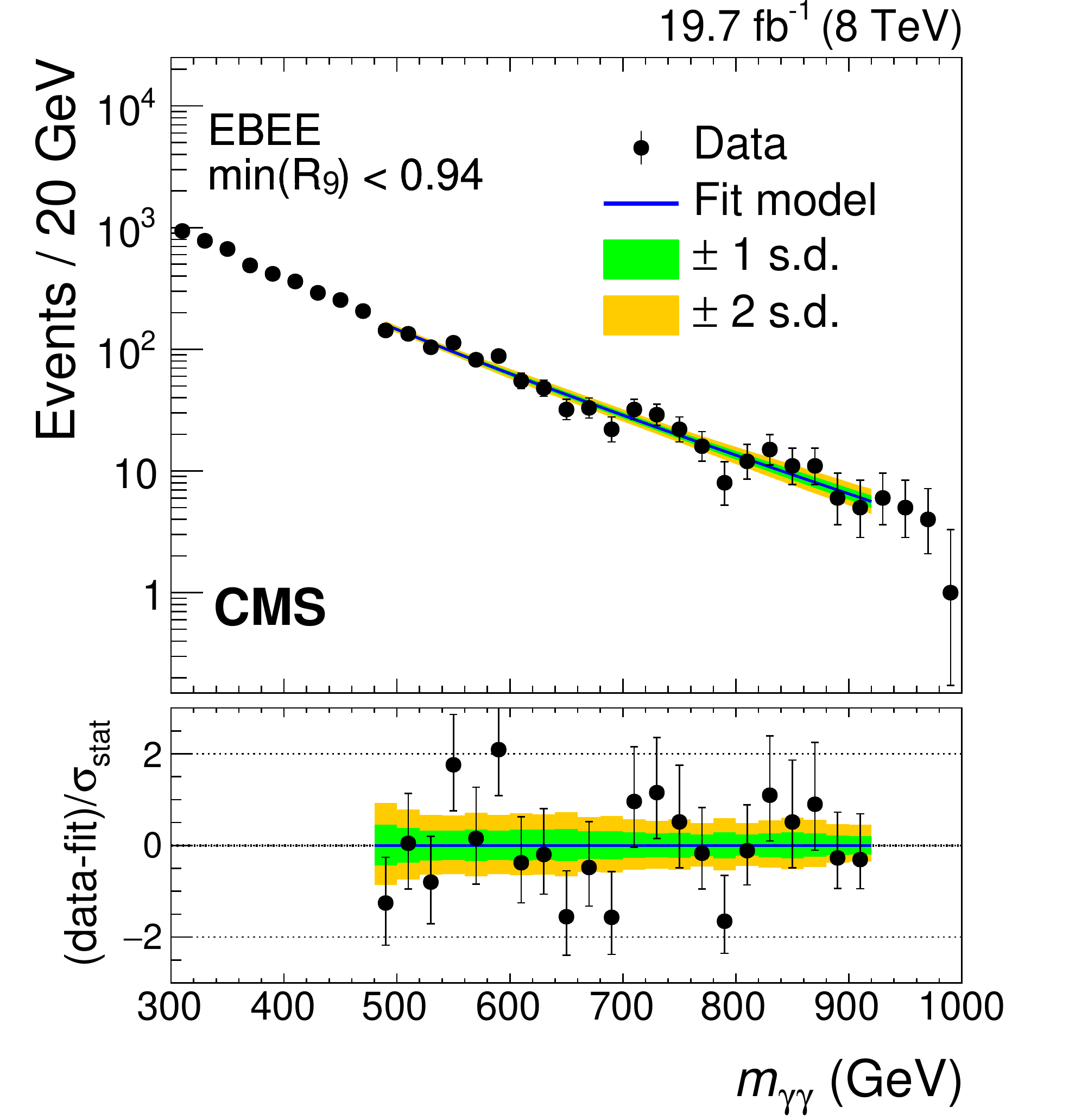} \\
    \caption{
      Observed diphoton invariant mass \mgg spectra for the event categories
      used in the analysis of the 8\TeV data for resonance mass $\mX\leq 850\GeV$:
      (upper row) $\min(\Rnine)>0.94$, (lower row) $\min(\Rnine)\leq0.94$;
      (left column) both photons in the ECAL barrel detector;
      (right column) one photon in the ECAL barrel detector and the other
      in the ECAL endcap detector.
      The results of background-only parametric fits to the data corresponding to the fit
      range used for the $\mX = 750\GeV$ hypothesis test are also shown~\cite{CMS-dipho-8TeV}.
      The shaded regions show the 1 and 2 standard deviation uncertainty bands associated with
      the fit, and reflect the statistical uncertainty of the data.
      The lower panels show the difference between the data and fit, divided by the
      statistical uncertainty in the data points.
      \label{fig:fits_8TeV}
    }
\end{figure*}

The \mgg distributions of the events selected in the 13\TeV analysis are shown in
Fig.~\ref{fig:fits_13TeV}. The corresponding 8\TeV results used for the $\mX \leq 850\GeV$
search are shown in Fig.~\ref{fig:fits_8TeV}~\cite{CMS-dipho-8TeV}. The \mgg distributions
of 8\TeV events used
for the  $\mX>850\GeV$ search are available in \suppMaterial.

The results of the search are interpreted in the framework of a composite statistical
hypothesis test. For each signal hypothesis, a simultaneous unbinned extended maximum
likelihood fit to the \mgg
spectra observed in all categories is performed and the likelihood function  used to
construct the test statistic. The modified frequentist method~\cite{Junk1999,bib-cls} is
utilized
to set upper limits on the production of diphoton resonances, following the prescription
described in Ref.~\cite{ATLAS:1379837}. The compatibility of the observation with the
background-only hypothesis is evaluated by computing the background-only
$p$-value~\cite{ATLAS:1379837}, denoted $p_0$ in the following. Asymptotic
formulas~\cite{Cowan:2010js} are used in the  calculations.
The accuracy of the formulas in the estimation of limits and significance is studied for a
subset of the hypothesis tests and is found to be about 10\%.
Thus the upper limits on the production cross section times branching fraction
for the resonant production of two photons could be up to 10\% higher,
and the significance of an excess over the SM up to 10\% lower,
than the results presented below.

The shape of the \mgg signal distribution in the likelihood function
is given by the
convolution of the intrinsic shape, taken from the PYTHIA generator, with a function
characterizing the CMS detector response.
The normalization is a free parameter of the fit. The intrinsic shape is generated for
various \mX
values. The detector response is derived from a \PYTHIA sample including $\GEANT4$ modeling
using a coarser spacing in \mX, assuming a small intrinsic width, and incorporating
corrections derived from \Zee data.
The intrinsic width and detector response are interpolated to intermediate points using
the ``moment morphing'' technique of Ref.~\cite{Baak:2015}.
At 13\TeV, the signal mass resolution, defined as the ratio of the full width at half
maximum (FWHM) of the distribution, divided by 2.35, to the peak position, is
roughly 1.0 (1.5)\% for the EBEB (EBEE) categories.

The background \mgg spectra are described by parametric functions of \mgg. The
coefficients are obtained from a fit to the data events, and considered as unconstrained
nuisance parameters in the fit. In this manner, the description of the background is derived from
data.
For the 13\TeV data and for the 8\TeV data in the $\mX > 850\GeV$ search, a parametrization
of the form $f(\mgg) = \mgg^{a + b \log( \mgg )}$ is chosen, where $a$ and
$b$ are parameters determined independently for each of the five event categories: the four shown
in Fig.~\ref{fig:fits_13TeV} plus that of the 8\TeV $\mX>850\GeV$ search.
The validity of the procedure is tested, using simulated background samples, by examining the difference
between the true and predicted numbers of background events in 14 contiguous intervals in \mgg within
the search region.
For each interval, a sampling distribution of the pull variable is constructed using
pseudo-experiments with the same sample size as the data. Background-only fits are
performed on the pseudo-experiments using the same \mgg ranges employed in data. In each
region, the pull is defined as the difference between the true and estimated numbers of
events divided by the estimated statistical uncertainty.
If the absolute value $\abs{m}$ of the median of the sampling distribution exceeds 0.5 in any
interval, the statistical uncertainty in the predicted number of background events is
increased by an additional term, denoted the ``bias term'', which is parameterized as a
continuous function of \mgg. The bias term is tuned in such a manner that the sampling
distribution of a pull variable that includes the bias term yields $\abs{m}<0.5$ for
all intervals.
The additional uncertainty is then included in the likelihood function by adding
to the background model a component having the same shape as the signal, with a
normalization coefficient distributed as a Gaussian of mean zero and width equal to the
integral of the bias term over the FWHM of the tested signal shape. The inclusion of the
additional component, whose magnitude is comparable to the 1 standard deviation band shown in Fig.~\ref{fig:fits_13TeV}, has the effect of avoiding falsely positive or negative tests that
could be induced by a mismodelling of the background shape, and it degrades the analysis
sensitivity by 5\% or less.

For the 8\TeV data in the $\mX \leq 850\GeV$ search, the background shape is parameterized as
$g(\mgg)=\mgg^{-c}\re^{-d \mgg}$, where $c$ and $d$ are parameters fit independently for
each event category of Fig.~\ref{fig:fits_8TeV}, and different \mgg intervals are used for
each \mX.
The intervals are chosen by comparing the results of the nominal
parameterization with those obtained using alternative parameterizations
of the background, with the intervals determined to minimize differences
in the predicted background yields~\cite{CMS-dipho-8TeV}.
The method used for 13\TeV and the one of Ref.~\cite{CMS-dipho-8TeV} yield similar
levels of uncertainty in the background estimation.
The latter approach, however, is not easily applicable when only a small number of events
populate the $\mgg>\mX$ region, which is why this approach is not adopted for the
13\TeV analysis or for the 8\TeV search with $\mX>850\GeV$.

We evaluate systematic uncertainties in the signal model predictions. For the 8\TeV data,
these are discussed in Ref.~\cite{CMS-dipho-8TeV}.
For the 13\TeV analysis they are as follows. For \bonNoUnit (\boffNoUnit)\bUnit, a 2.7\,(12)\%
uncertainty is due
to the
limited knowledge of the total integrated luminosity~\cite{CMS-PAS-LUM-15-001}. An 8\,(16)\% uncertainty is attributed to the selection efficiency and a 6\,(6)\% uncertainty to
the PDFs.
An uncertainty of 1\% is assigned to the absolute photon energy scale,
with an additional 1\% to account for possible differences
between the energy scales of the 3.8 and 0\unit{T} samples.
An uncertainty in the signal mass resolution is assessed by varying the photon energy
resolution corrections derived from \Zee events by ${\pm}0.5\%$.
Energy resolution uncertainties are taken to be uncorrelated between the 8 and
13\TeV data, while a linear correlation of 0.5 is assumed for the energy scale.
Taking the value of the linear correlation to be 0 or 1 leads to negligible changes in the
results. Other systematic uncertainties are taken to be uncorrelated between the two
data sets, except for the one associated with the PDFs, which is taken to be fully correlated.

The ratio of the 8\TeV to the 13\TeV production rates is determined from
simulation and is held constant in the fit.
For the scalar (RS graviton) resonance, this ratio decreases from 0.27 (0.29) at
$\mX=500\GeV$ to 0.03 (0.04) at $\mX=4\TeV$ and equals 0.22 (0.24) for $\mX=750\GeV$.
The uncertainty in this ratio, determined by varying the PDFs, is found to have a
negligible impact on the results and is therefore ignored.

\begin{figure}[hbt]
    \centering
    \includegraphics[width=\cmsFigMedWidth]{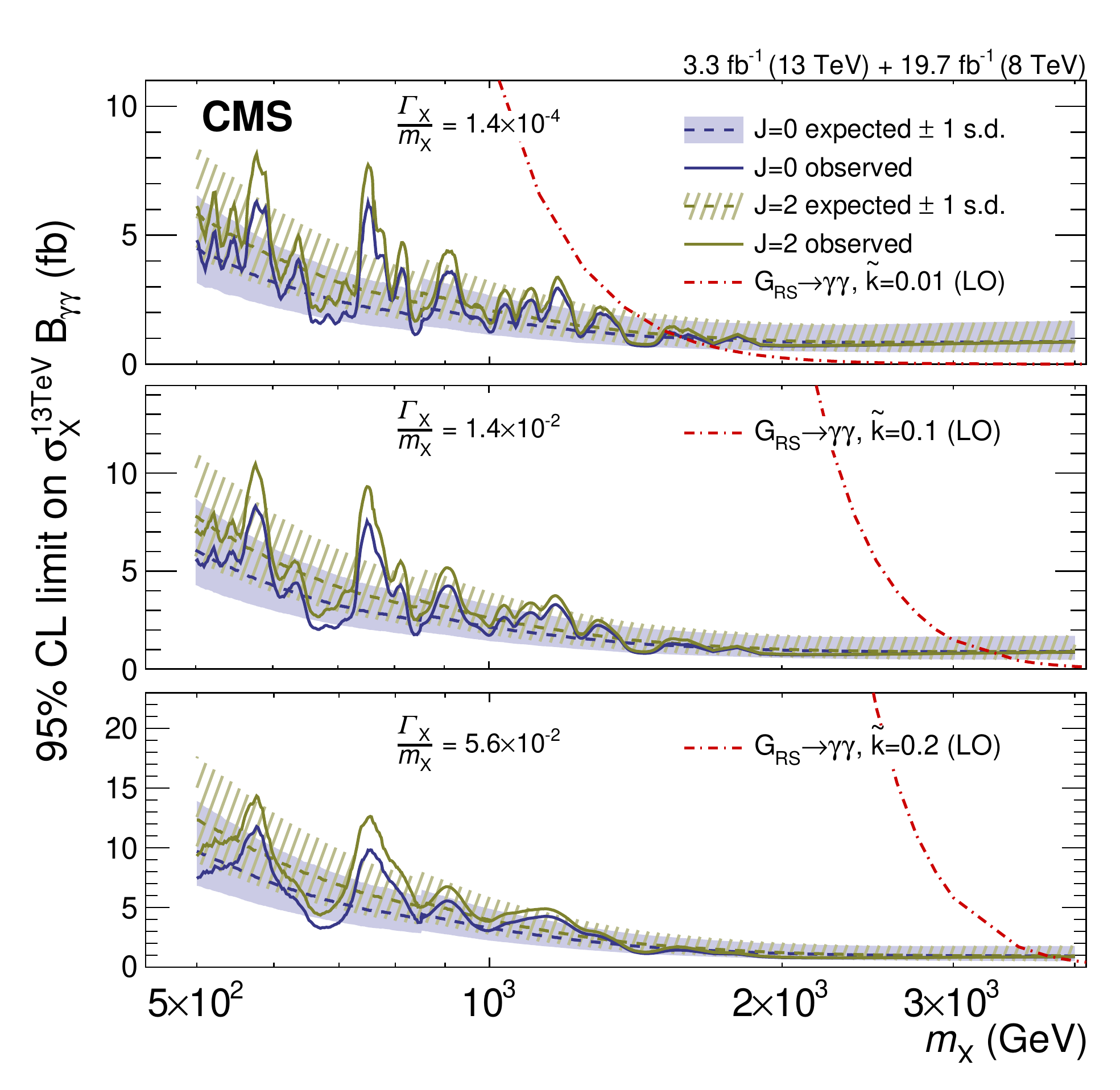}
    \caption{
      The 95\% \CL upper limits on the production of diphoton resonances as a
      function of the resonance mass \mX, from the combined analysis of the
      8 and 13\TeV data.
      The 8\TeV results are scaled by the ratio of the 8 to 13\TeV cross sections.
      The blue-grey (darker) curves and the green (lighter) ones correspond to the scalar and
      RS graviton signals, respectively. Solid (dashed) curves represent the observed
      (median expected) exclusion limit. The expected results are shown with their 1
      standard deviation dispersion bands.
      The leading-order \RS graviton production cross section is shown by the red dot-dashed
      curves.
      The results are shown for (upper) a narrow, (middle) an intermediate-width, and
      (lower) a broad resonance, with the value of the width \GammaOm, relative to the
      mass, indicated in the legend of each plot.
      \label{fig:comb_limits}
    }
\end{figure}

The median expected and observed 95\% confidence level (\CL) exclusion limits on the
product of the 13\TeV signal production cross section and  decay branching fraction,
$\sigma^{13\TeV}_{X} \mathcal{B}_{\gaga}$, are presented in Fig.~\ref{fig:comb_limits}
for the combined analysis.
The upper (lower) plot shows the results for a narrow (broad) resonance width, $\GammaOm =
1.4\times 10^{-4}$ ($5.6\times 10^{-2}$).  The results for $\GammaOm = 1.4\times 10^{-2}$ are shown
  in the middle plot.
The  blue-grey (darker) and green (lighter) solid curves indicate the observed limits for
a scalar resonance and an RS graviton.  The corresponding dashed curves show the expected
limits, with their one standard deviation intervals.
Using the LO cross sections from \PYTHIA\,8.2, \RS gravitons with masses below 1.6,
3.3, and 3.8\TeV are excluded for $\ktild=0.01$, $0.1$, and $0.2$, respectively,
corresponding to $\GammaOm=1.4\times10^{-4}$, $1.4\times10^{-2}$, and $5.6\times10^{-2}$.

\begin{figure}[hbt]
    \centering
    \includegraphics[width=\cmsFigMedWidth]{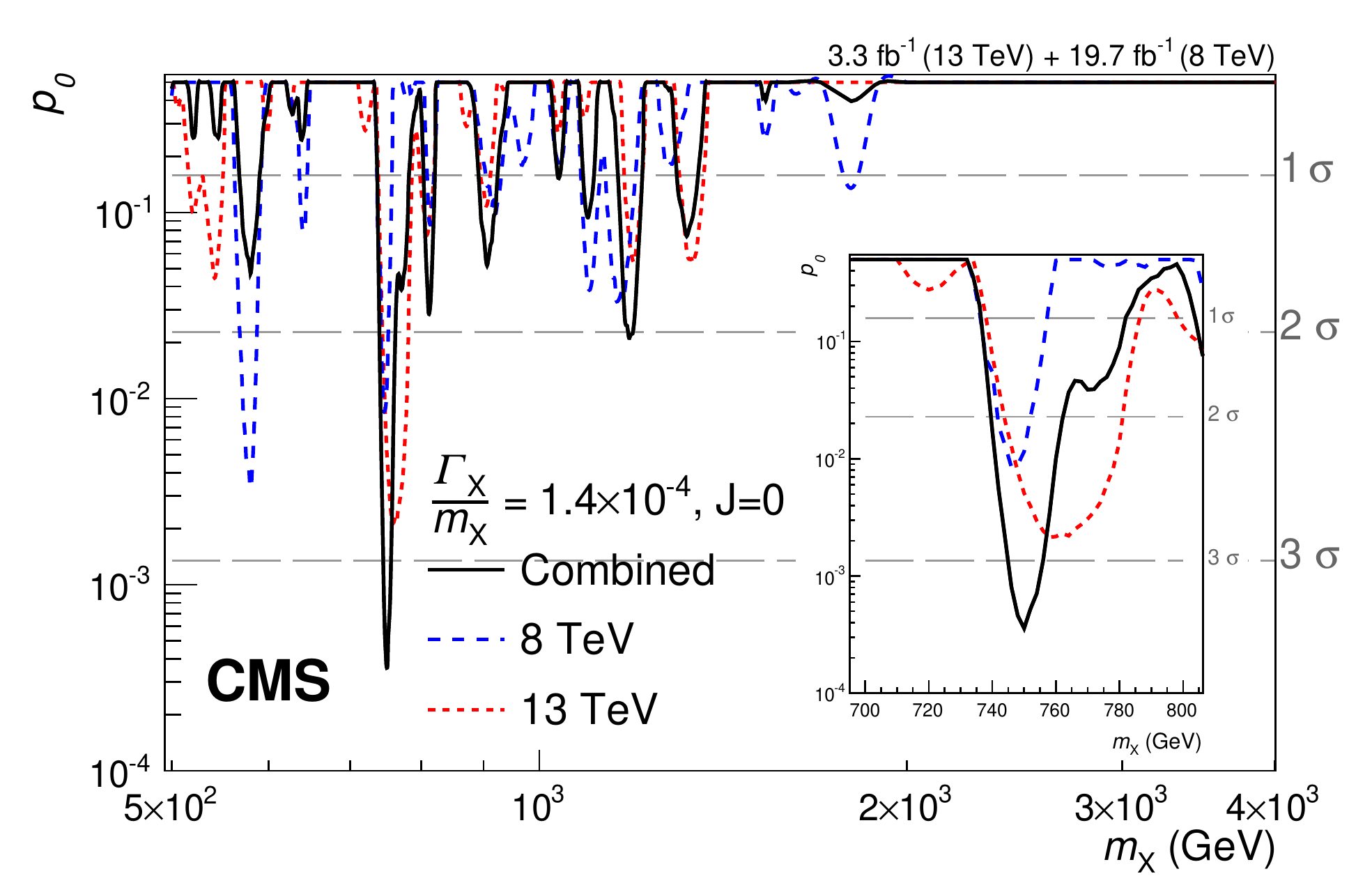}
    \caption{
      Observed background-only $p$-values for narrow-width scalar resonances as a function of
      the resonance mass \mX, from the combined analysis of the 8 and 13\TeV data.
      The results for the separate 8 and 13\TeV data sets are also shown.
      The inset shows an expanded region around $\mX=750\GeV$.
      \label{fig:comb_pvalues}
    }
\end{figure}

The observed value of $p_0$ as a function of \mX is shown in
Fig.~\ref{fig:comb_pvalues} for the scalar narrow-width hypothesis ($\GammaOm=1.4\times 10^{-4}$).
The largest excess, observed for $\mX \approx 750\GeV$, has a local significance of
approximately 3.4 standard deviations. Similar values are obtained for the two spin
hypotheses, while lower values of the local significance are obtained for wider signal
hypotheses. For $\GammaOm=5.6\times10^{-2}$ a local significance of 2.3 standard
deviations is estimated.

Trial factors associated with the test of several mass hypotheses are estimated for fixed
width and spin assumptions by counting the number of times the value of $p_0$ observed in
data
crosses the level corresponding to 0.5 standard deviations and applying the asymptotic
formulas of Ref.~\cite{Gross:2010qma}, where a trial factor refers to the ratio of the
probability to
observe an excess at a given \mX value to the probability to observe it anywhere in the
examined \mX range.
To account for the different width and spin hypotheses tested, a correction factor is
estimated using the 13\TeV event categories,  as follows.  A sampling
distribution of the minimum value of $p_0$ is generated from an ensemble of
background-only pseudo-experiments, testing for all examined spin, width, and mass
hypotheses.  The correction factor is given by the ratio of the trial factors obtained
varying only the signal mass to those obtained also varying the width and spin.
A global significance for the 750\GeV excess, taking into account the effect of testing all
the signal hypotheses considered, is thereby estimated to be approximately 1.6 standard
deviations.
The estimated global significance increases by about 5\% if the spin hypothesis is not
varied and by an additional 5\% if only narrow-width signal hypotheses are
considered. A statistical uncertainty of roughly 10\% in the estimated global significance
is associated with the counting of $p_0$ crossings in data.

The excess is primarily due to events in which both photons are in the ECAL
barrel. The shape of the associated ECAL clusters is in agreement with the expectation for
high-\pt prompt photons. In particular, the \Rnine value exceeds 0.94
for more than 80\% of the photon pair candidates in the 13\TeV data in
the region corresponding to the excess,
i.e., the showers are compact,
with lateral shapes like those of unconverted photons at lower energy,
in agreement with the expectation for a sample
of prompt high energy photon pairs.
Within the limited statistical precision currently available, the kinematic distributions
of the diphoton candidates in the \mgg region corresponding to the largest excess, as well as
the multiplicity and kinematic distributions of the hadronic jets reconstructed in the
same events, do not exhibit significant deviations from the distributions expected for SM
processes.

In summary, a search for the resonant production of high-mass photon pairs is presented.
The analysis is based on 19.7 and \mylumi of proton-proton collisions collected at
$\sqrts=8$ and  13\TeV, respectively,  by the CMS experiment.
Limits on the production cross section of scalar resonances and Randall--Sundrum gravitons
for resonance
masses ${0.5<\mX<4\TeV}$ and relative widths $1.4\times 10^{-4} < \GammaOm <
5.6\times10^{-2}$ are determined.
Using leading-order cross sections for \RS graviton production, \RS gravitons with masses
below about 1.6, 3.3, and 3.8\TeV are excluded at 95\% confidence level  for
$\ktild=0.01$, $0.1$, and $0.2$, respectively, corresponding to
$\GammaOm=1.4\times10^{-4}$, $1.4\times10^{-2}$,  and $5.6\times10^{-2}$.
A modest excess of events over the background-only hypothesis is observed for
$\mX \approx 750\GeV$. The local  $p$-value under the narrow-width hypothesis of
$\GammaOm=1.4 \times 10^{-4} $ is 3.4 standard deviations.
At $\mX=750\GeV$, the 8 and 13\TeV data contribute with similar weights to the
combined result.
The significance of the excess is reduced to about 1.6 standard deviations
once the effect of searching under multiple signal hypotheses is taken
into account. More data are required to determine the origin of this excess.
A similar analysis is presented by the ATLAS Collaboration~\cite{atlas-gg}.

\begin{acknowledgments}
We congratulate our colleagues in the CERN accelerator departments for the excellent performance of the LHC and thank the technical and administrative staffs at CERN and at other CMS institutes for their contributions to the success of the CMS effort. In addition, we gratefully acknowledge the computing centres and personnel of the Worldwide LHC Computing Grid for delivering so effectively the computing infrastructure essential to our analyses. Finally, we acknowledge the enduring support for the construction and operation of the LHC and the CMS detector provided by the following funding agencies: BMWFW and FWF (Austria); FNRS and FWO (Belgium); CNPq, CAPES, FAPERJ, and FAPESP (Brazil); MES (Bulgaria); CERN; CAS, MoST, and NSFC (China); COLCIENCIAS (Colombia); MSES and CSF (Croatia); RPF (Cyprus); SENESCYT (Ecuador); MoER, ERC IUT and ERDF (Estonia); Academy of Finland, MEC, and HIP (Finland); CEA and CNRS/IN2P3 (France); BMBF, DFG, and HGF (Germany); GSRT (Greece); OTKA and NIH (Hungary); DAE and DST (India); IPM (Iran); SFI (Ireland); INFN (Italy); MSIP and NRF (Republic of Korea); LAS (Lithuania); MOE and UM (Malaysia); BUAP, CINVESTAV, CONACYT, LNS, SEP, and UASLP-FAI (Mexico); MBIE (New Zealand); PAEC (Pakistan); MSHE and NSC (Poland); FCT (Portugal); JINR (Dubna); MON, RosAtom, RAS and RFBR (Russia); MESTD (Serbia); SEIDI and CPAN (Spain); Swiss Funding Agencies (Switzerland); MST (Taipei); ThEPCenter, IPST, STAR and NSTDA (Thailand); TUBITAK and TAEK (Turkey); NASU and SFFR (Ukraine); STFC (United Kingdom); DOE and NSF (USA).
\end{acknowledgments}

\bibliography{auto_generated}
\numberwithin{figure}{section}
\ifthenelse{\boolean{cms@external}}{}{
\clearpage
\appendix
\section{Supplemental material
  \label{app:supp_material}}

\input{supp_material}

}
\cleardoublepage \section{The CMS Collaboration \label{app:collab}}\begin{sloppypar}\hyphenpenalty=5000\widowpenalty=500\clubpenalty=5000\input{EXO-16-018-authorlist.tex}\end{sloppypar}
\end{document}

%% file: supp_material.tex
\begin{figure}[hbt]
    \centering
    \includegraphics[width=\cmsFigWidth]{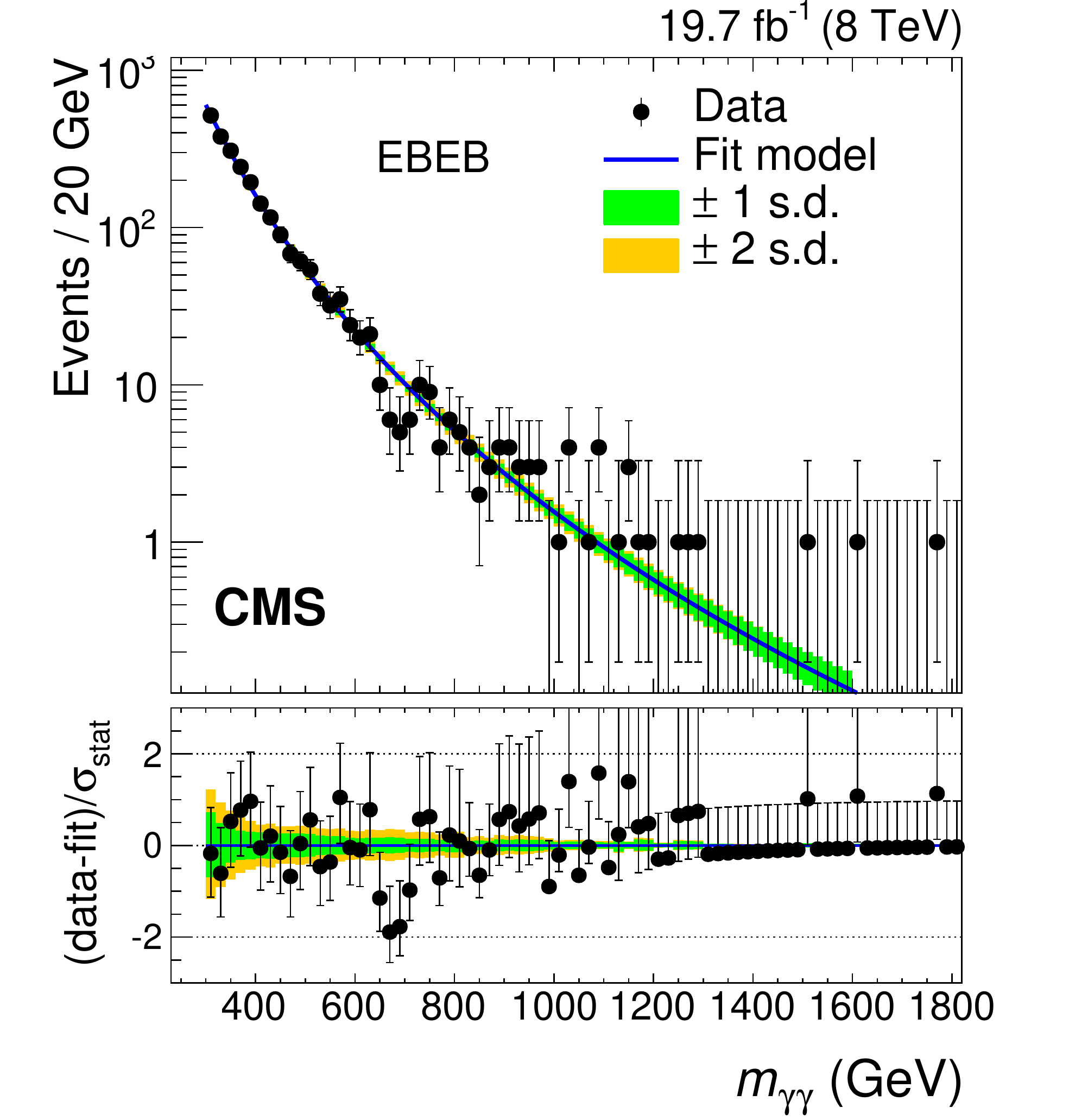}
    \caption{
      Observed diphoton invariant mass \mgg spectra for the event categories
      used in the analysis of the 8\TeV data for the $\mX>850\GeV$ search.
      No event with $\mgg>1800\GeV$ is selected in the analysis.
      The results of a likelihood fit to
      the background-only hypothesis are also shown.
      The lower panel shows the difference between the data and fit, divided by the
      statistical uncertainty in the data points.
    }
\end{figure}

\begin{figure}[hbtp]
    \centering
    \includegraphics[width=\cmsFigMedWidth]{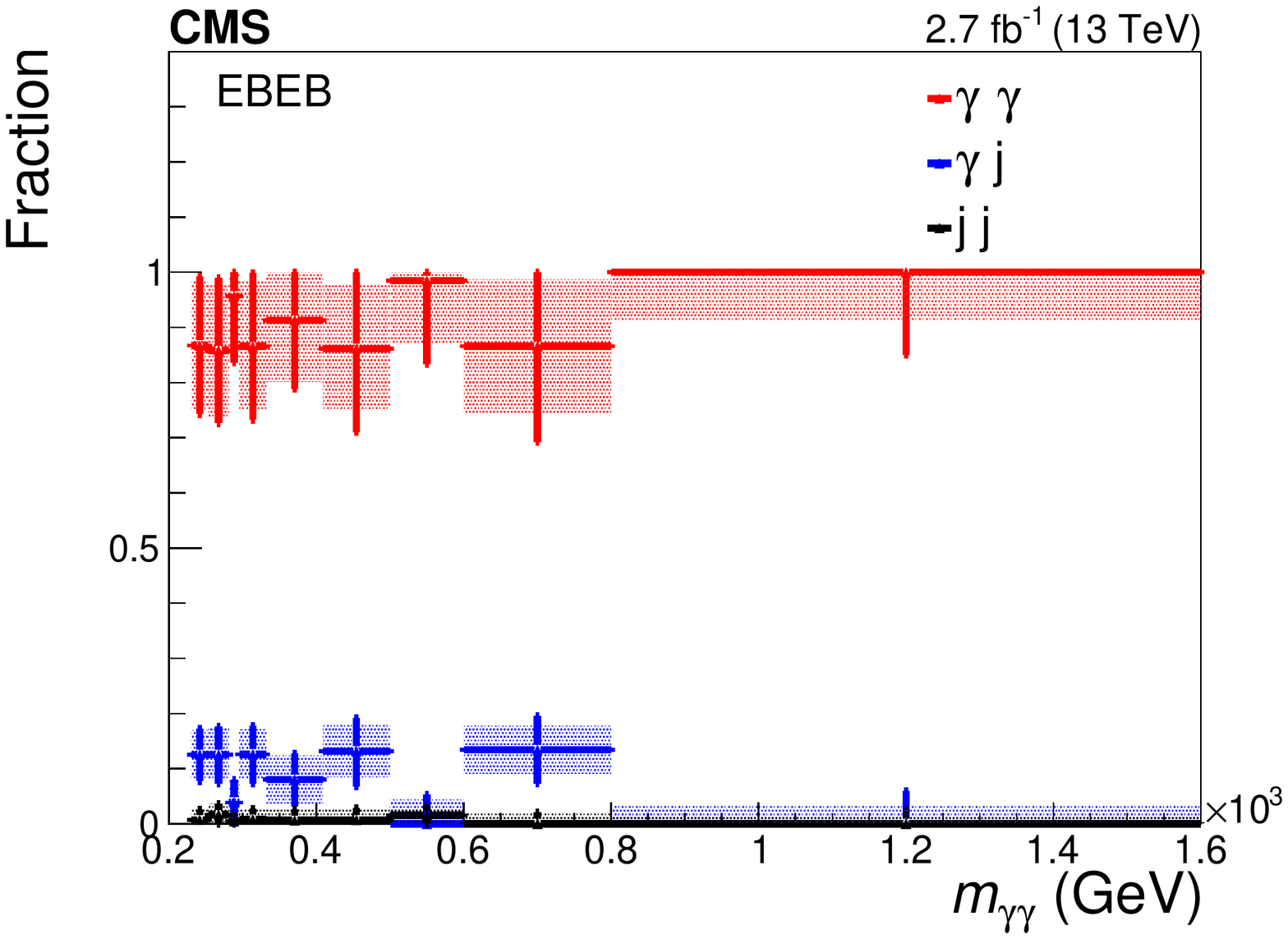}
    \includegraphics[width=\cmsFigMedWidth]{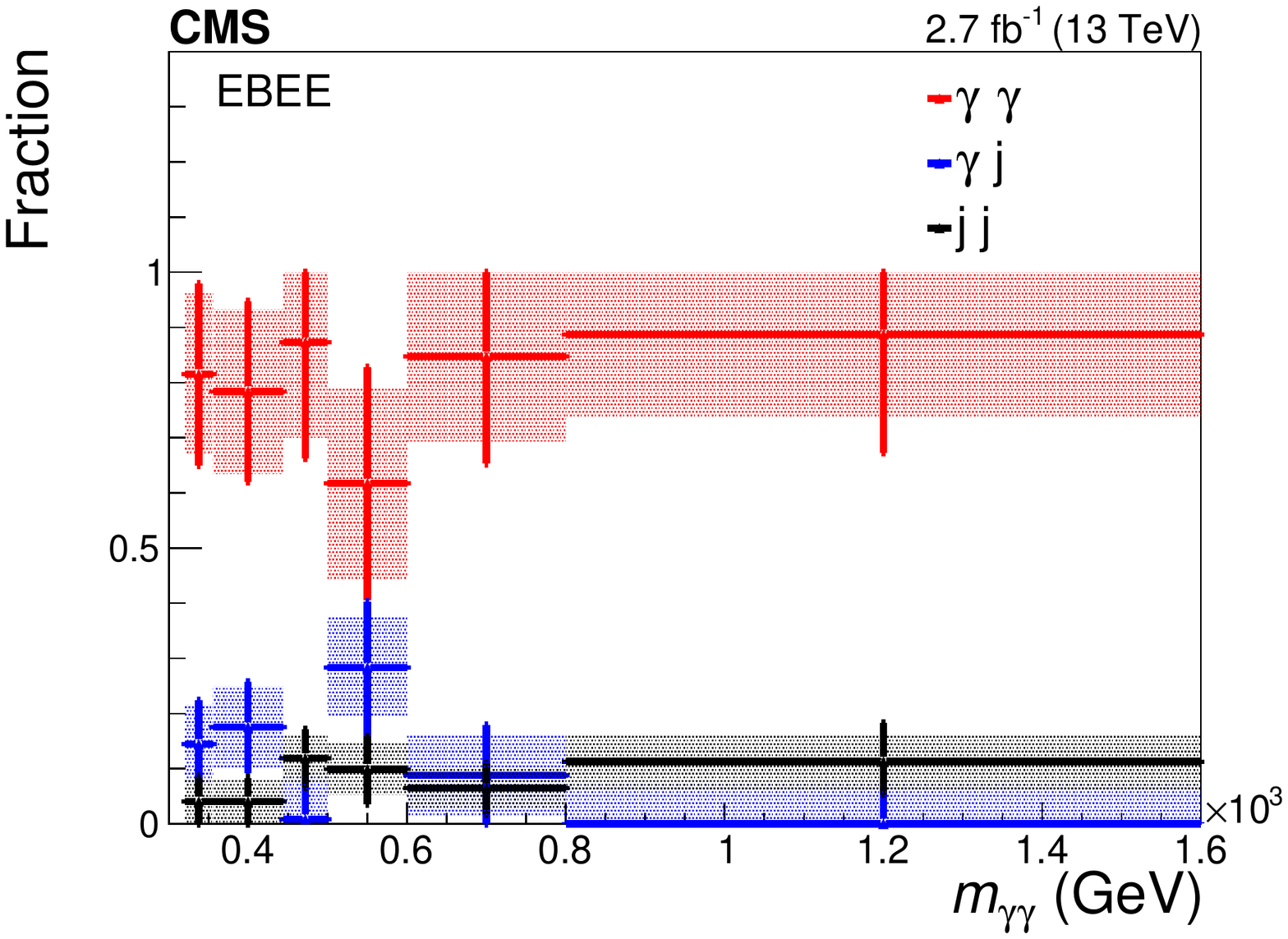}
    \caption{
      Measured composition of the background for the 13\TeV analysis at 3.8\unit{T} for
      the (\cmsLeft) EBEB and (\cmsRight) EBEE categories. The method described in
      Ref.~\cite{smdipho_7TeV} was used to obtain these results.
      The background corresponds to the direct production of two photons ($\gamma\gamma$),
      the production of $\gamma\,$+$\,$jets events ($\gamma j$),
      and the production of multijet events ($jj$).
      The shaded error boxes represent the systematic uncertainties associated with the
      measurement, while the error bars represent the total uncertainties, obtained adding
      in quadrature statistical and systematic uncertainties.
    }
\end{figure}

\begin{figure}[hbt]
    \centering
    \includegraphics[width=\cmsFigMedWidth]{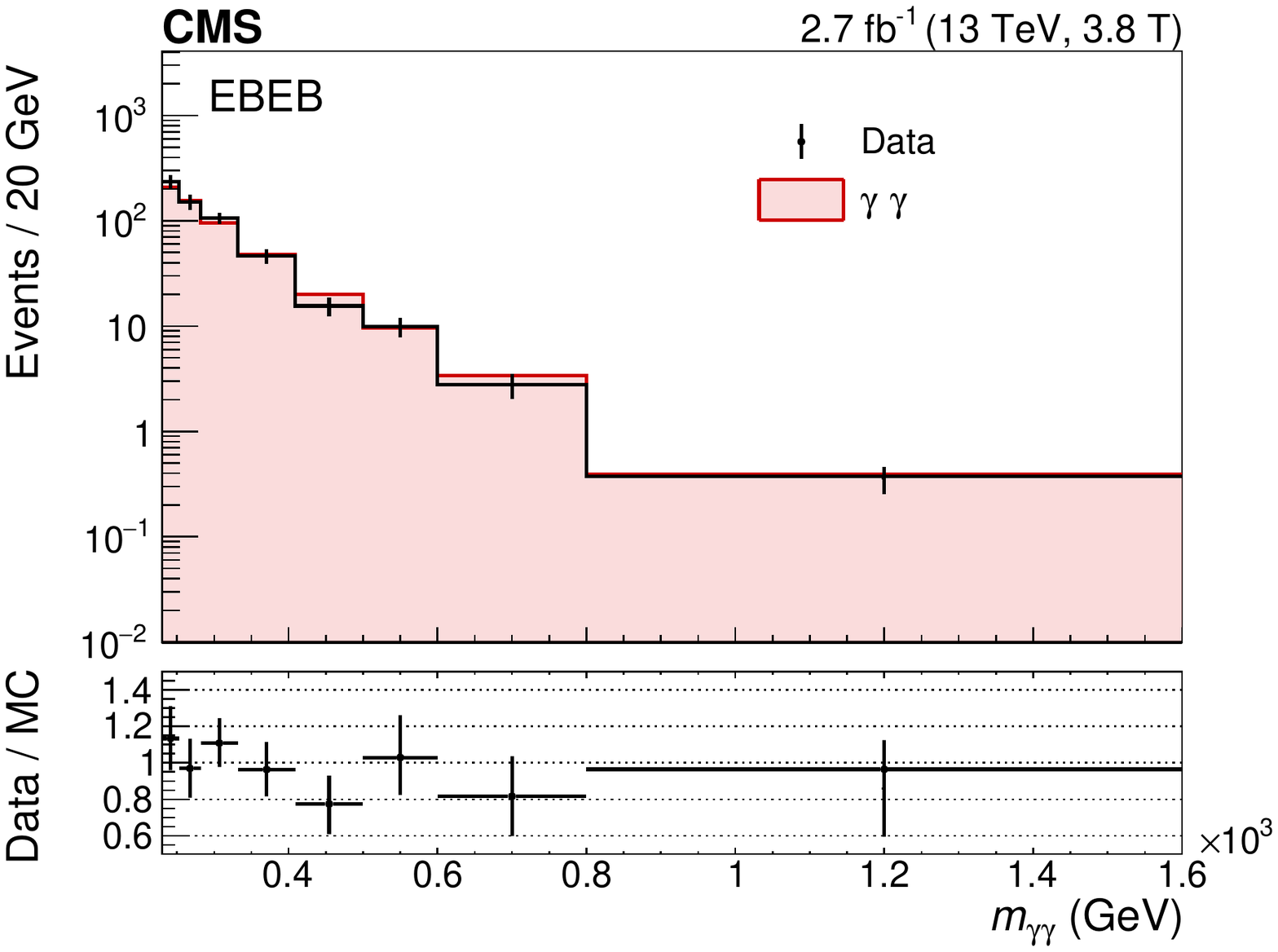}
    \includegraphics[width=\cmsFigMedWidth]{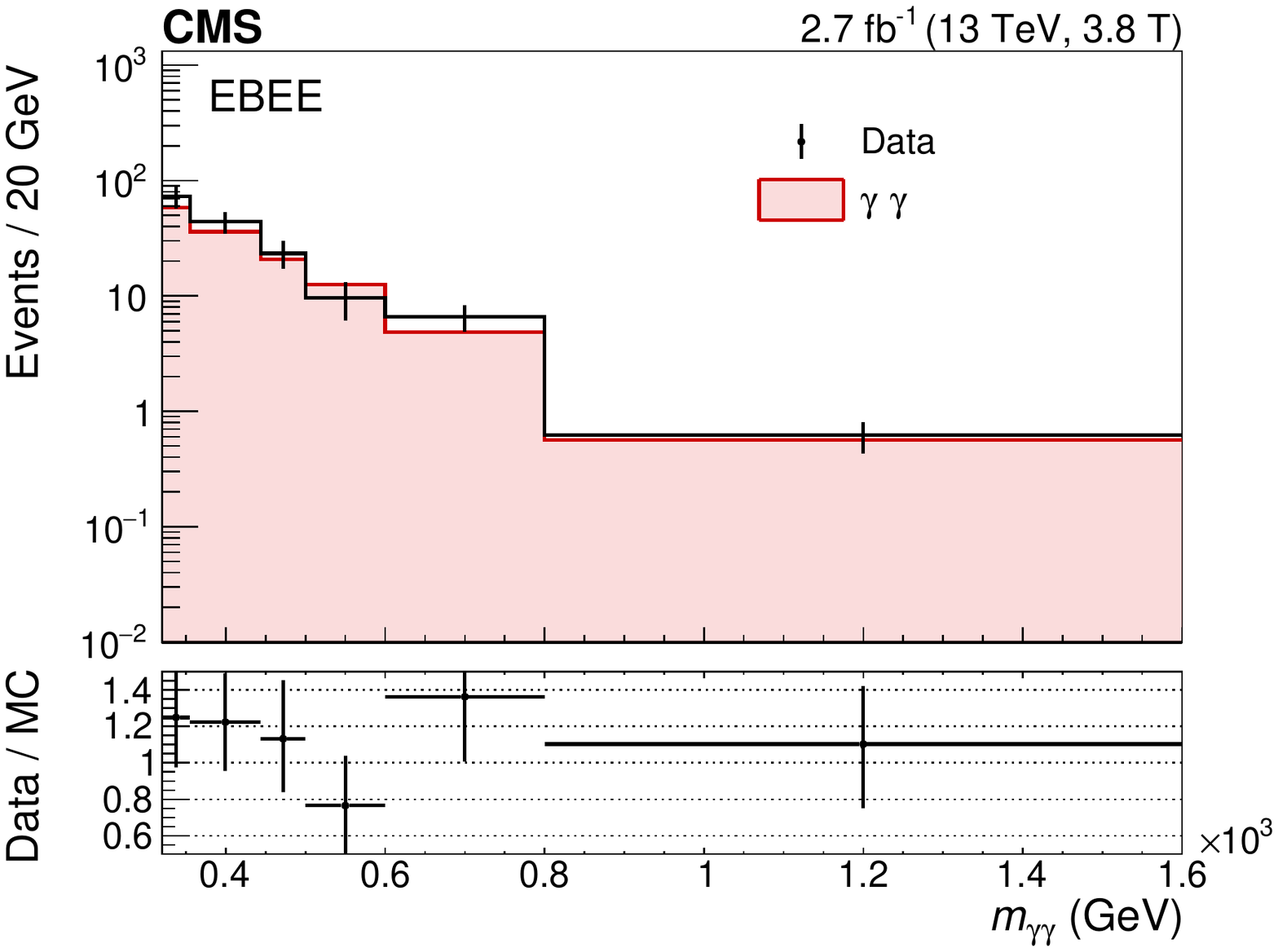}
    \caption{
      Comparison between the measured and the predicted invariant mass spectrum of the
      nonresonant
      SM $\gamma\gamma$ background for the 13\TeV analysis at 3.8\unit{T} for
      the
      (\cmsLeft) EBEB and (\cmsRight) EBEE categories.
The $\gamma+$jets and multijet  background components are subtracted in data,
using the method described in Ref.~\cite{smdipho_7TeV}.
The predicted background is obtained correcting the distribution of the events generated
with the  $\SHERPA\,2.1$~\cite{Gleisberg:2008ta} generator
(where the CMS detector response was simulated using the
\GEANTfour package~\cite{Agostinelli:2002hh})
to match the predictions
obtained with the $2\gamma$ NNLO program~\cite{Catani:2011qz}.
}
\end{figure}

\begin{figure}[hbt]
    \centering
    \includegraphics[width=\cmsFigMedWidth]{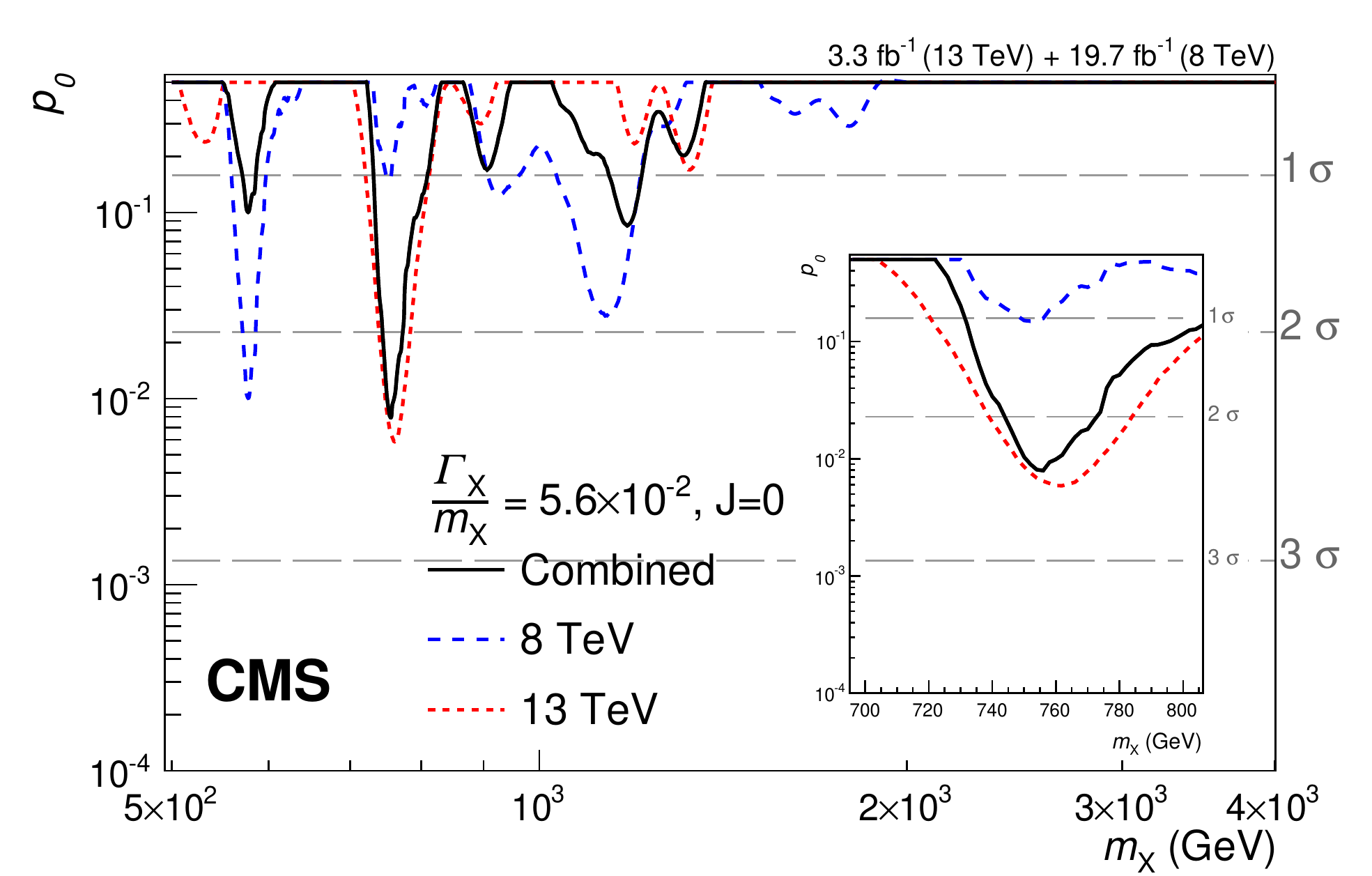}
    \caption{
      Observed background-only $p$-values for spin-0 resonances with $\GammaOm=5.6\times
      10^{-2}$ as a function of the resonance mass \mX from the combined analysis of the 8
      and 13\TeV data.
      The results for the 8 and 13\TeV data sets are also shown separately.
      \label{fig:extra_comb_pvalues}
    }
\end{figure}

\begin{figure}[hbtp]
    \centering
    \includegraphics[width=\cmsFigWidth]{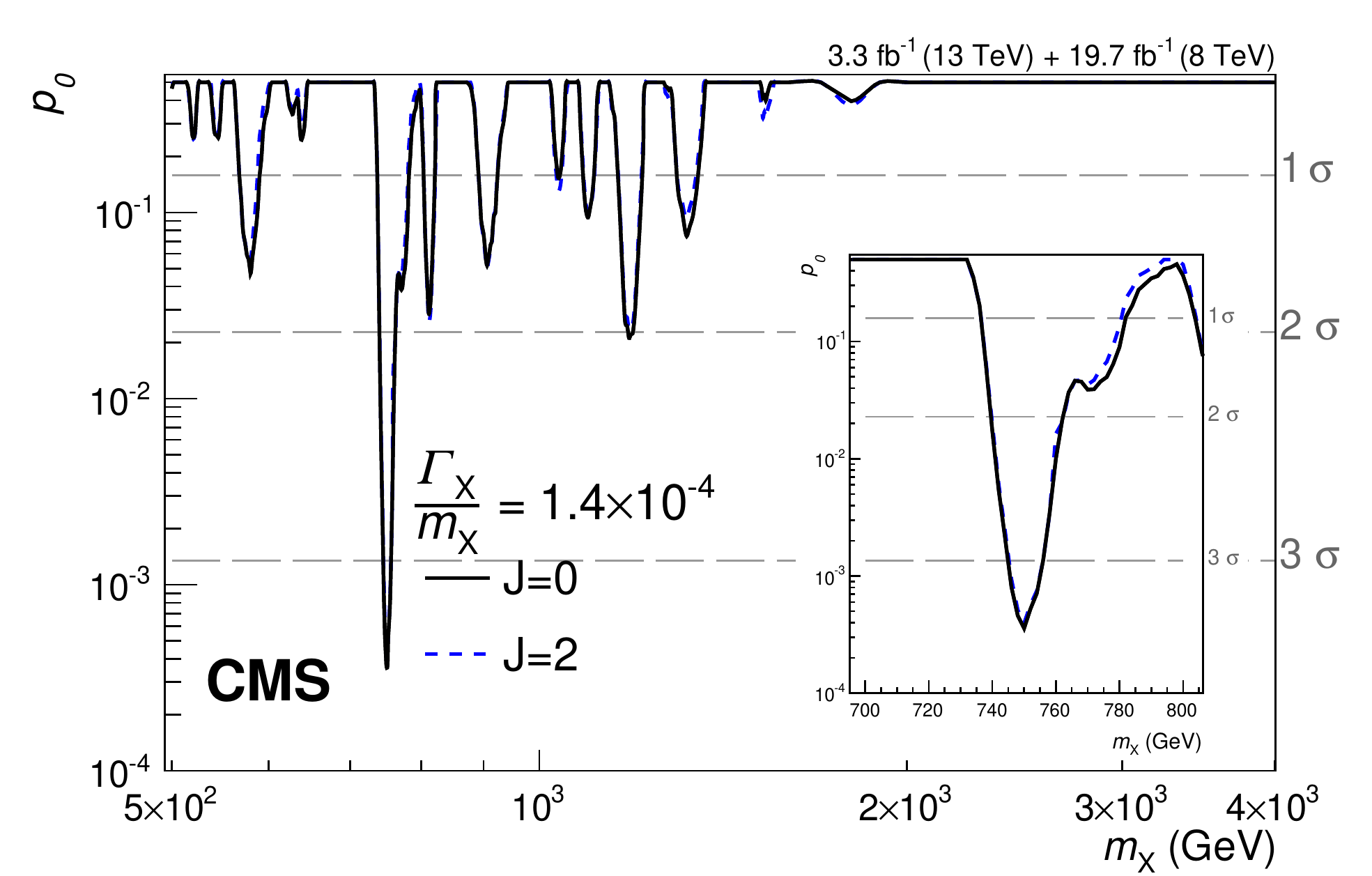}\\
    \includegraphics[width=\cmsFigWidth]{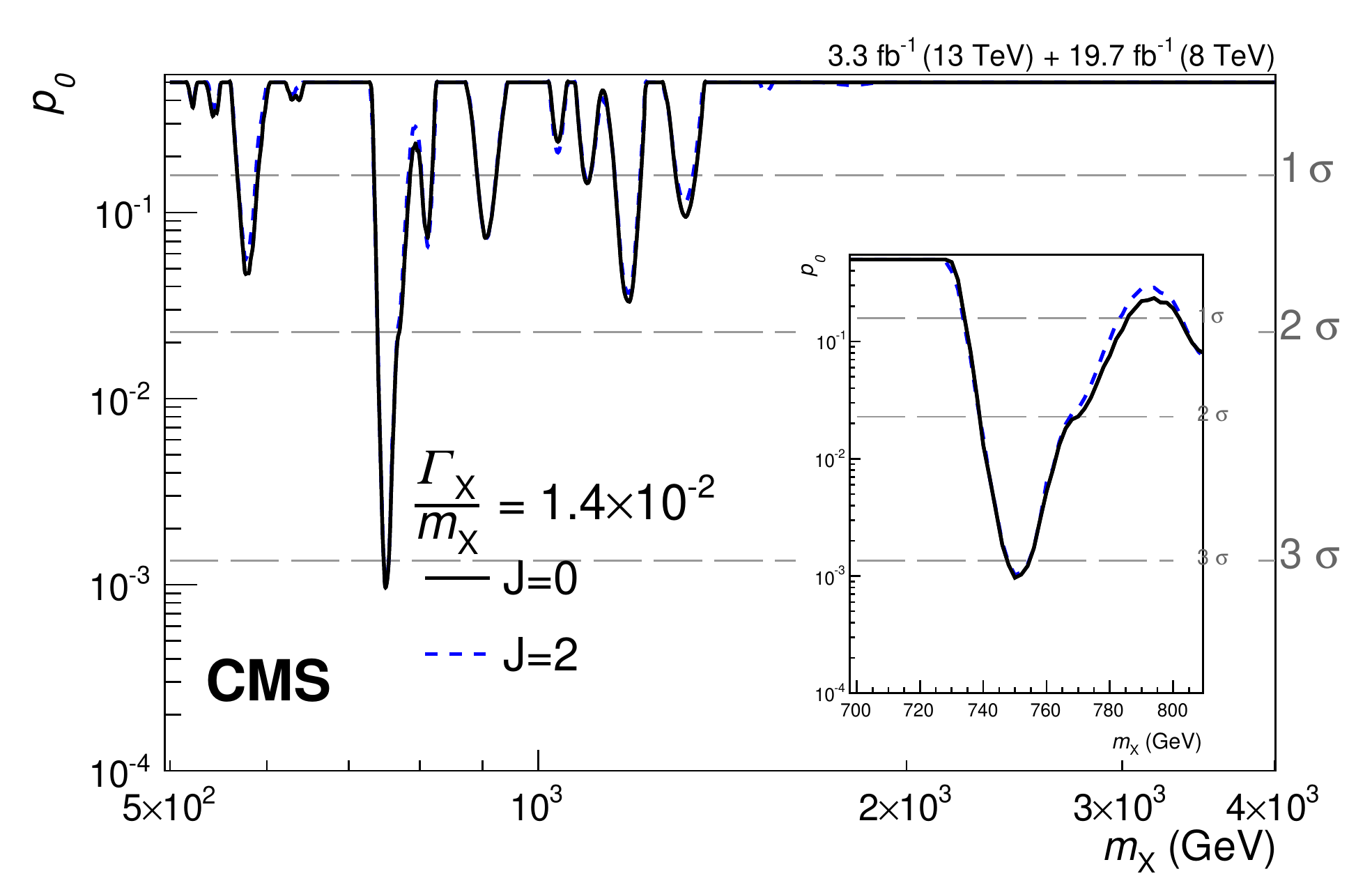} \\
    \includegraphics[width=\cmsFigWidth]{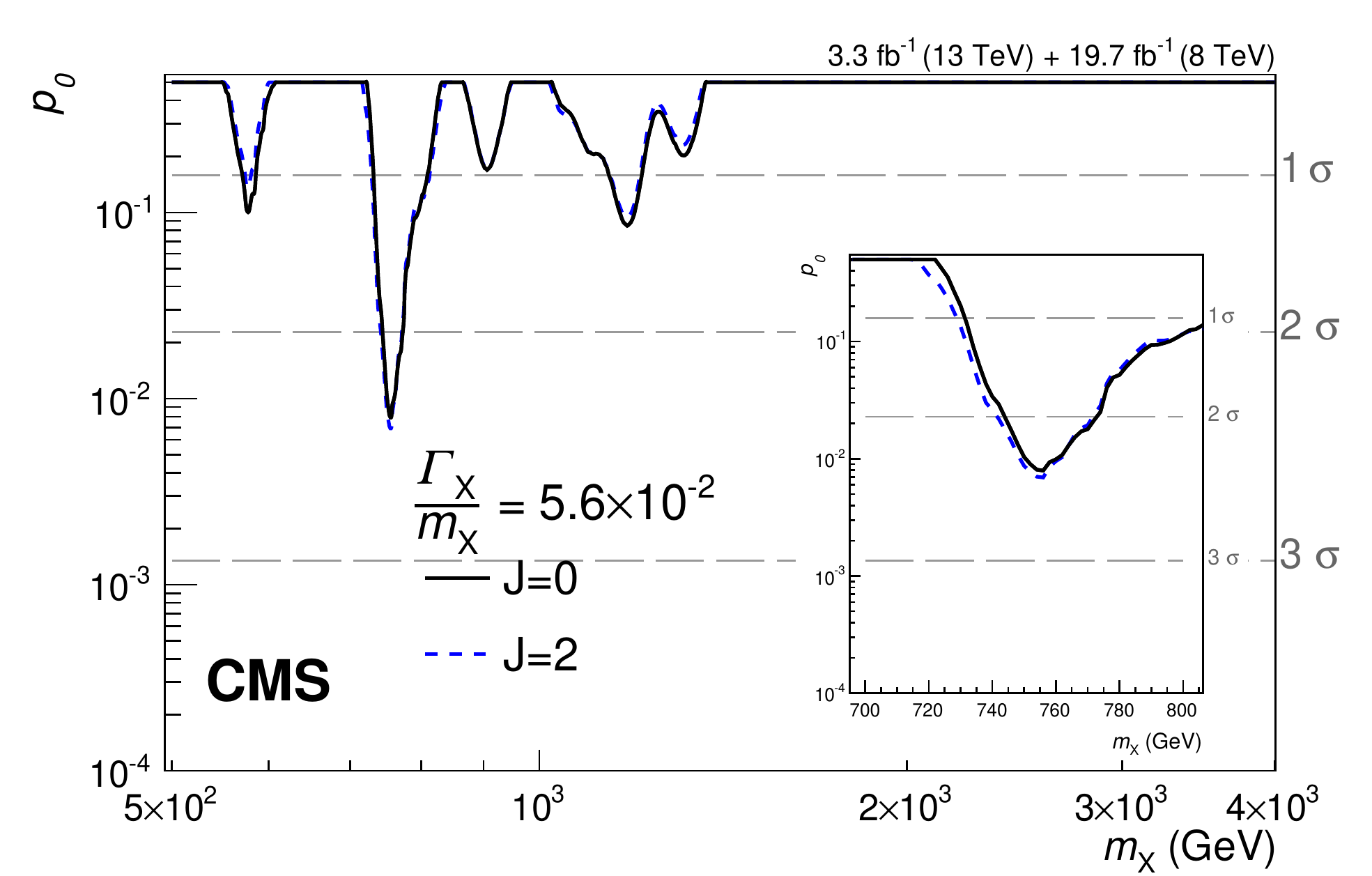}
    \caption{
      Observed background-only $p$-values as a function of the resonance mass \mX from the
      combined analysis of the 8 and 13\TeV data. Three width hypotheses are shown: (\cmsLeft)
      $\GammaOm = 1.4 \times 10^{-4}$; (middle) $\GammaOm = 1.4 \times 10^{-2}$; (\cmsRight)
      $\GammaOm = 5.6 \times 10^{-2}$.
      In each plot, the results obtained under the RS graviton and scalar hypotheses are shown.
      \label{fig:extra_comb_pvalues_2}
    }
\end{figure}

\begin{figure}[thb]
    \centering
    \includegraphics[width=\cmsFigWidth]{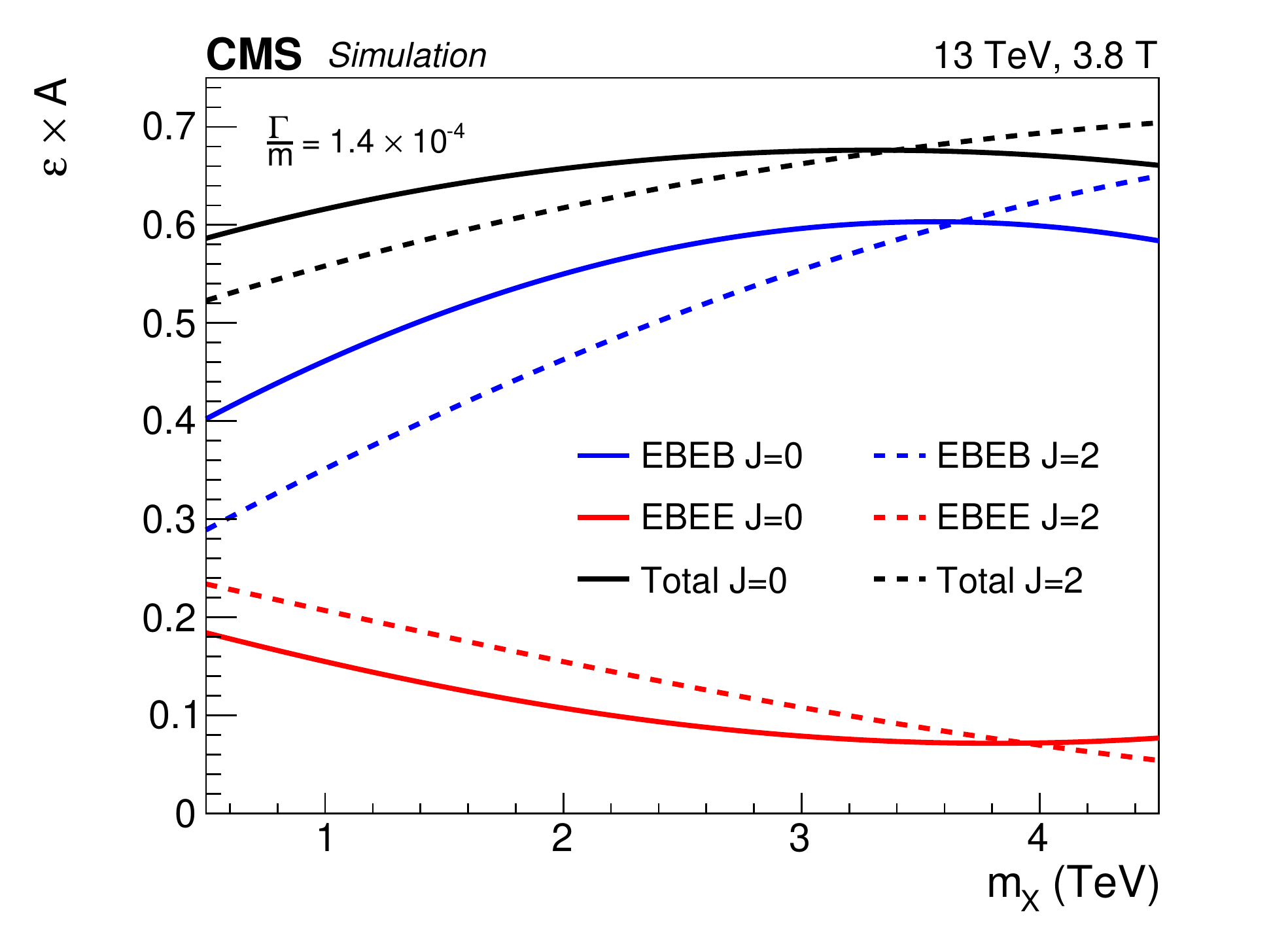}
    \includegraphics[width=\cmsFigWidth]{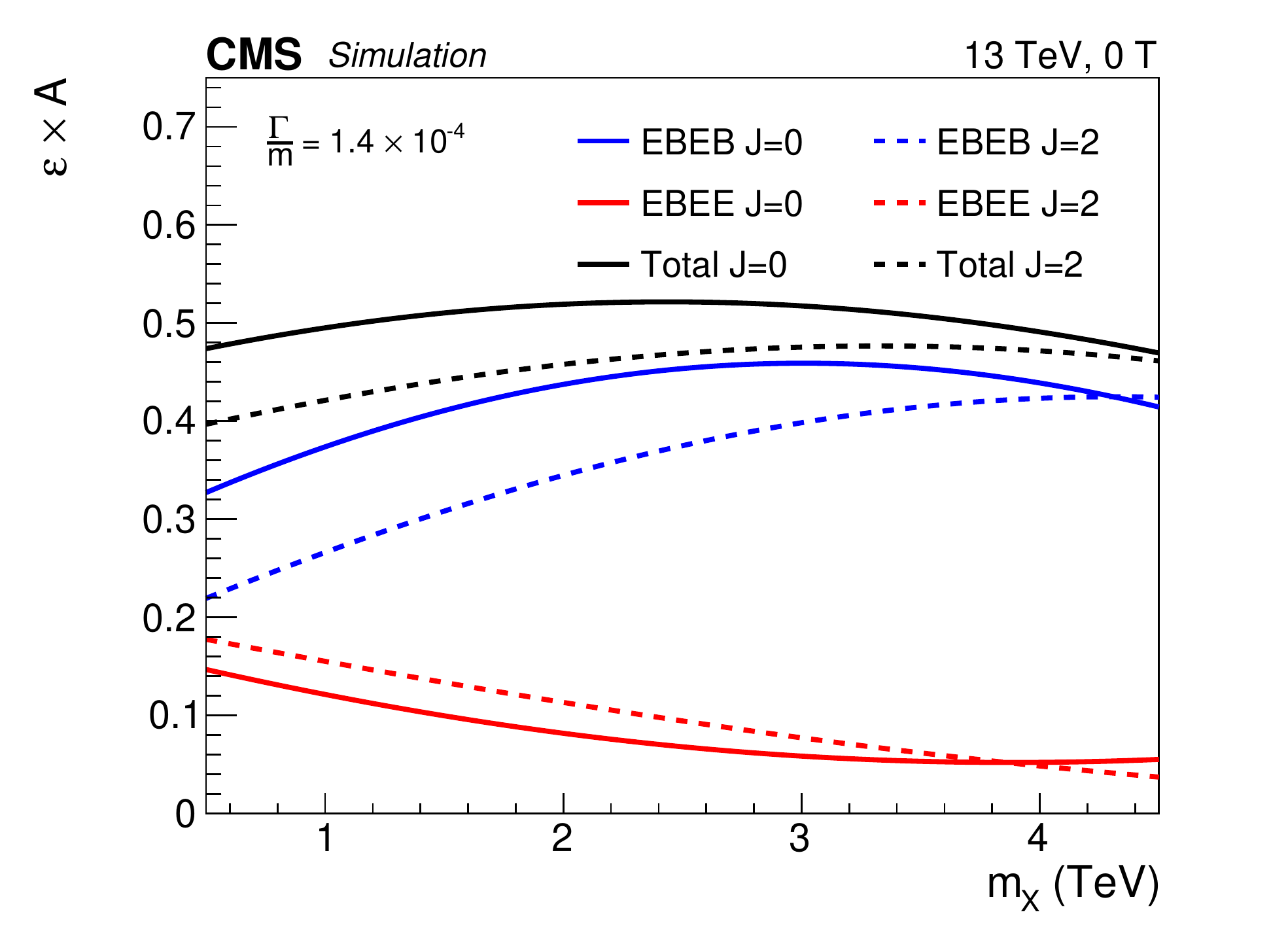}
    \caption{
      Fraction of events selected by the analysis categories for
      $0.5 < \mX < 4.5\TeV$ and $\GammaOm = 1.4\times 10^{-4}$.
      Curves for both spin-0 and \RS graviton resonances are shown, in the \cmsLeft plot for the
      \bon sample and in the \cmsRight one for the \boff sample.
      \label{fig:eff_acc}
    }
\end{figure}

\begin{figure}[hbtp]
    \centering
    \includegraphics[width=\cmsFigWidth]{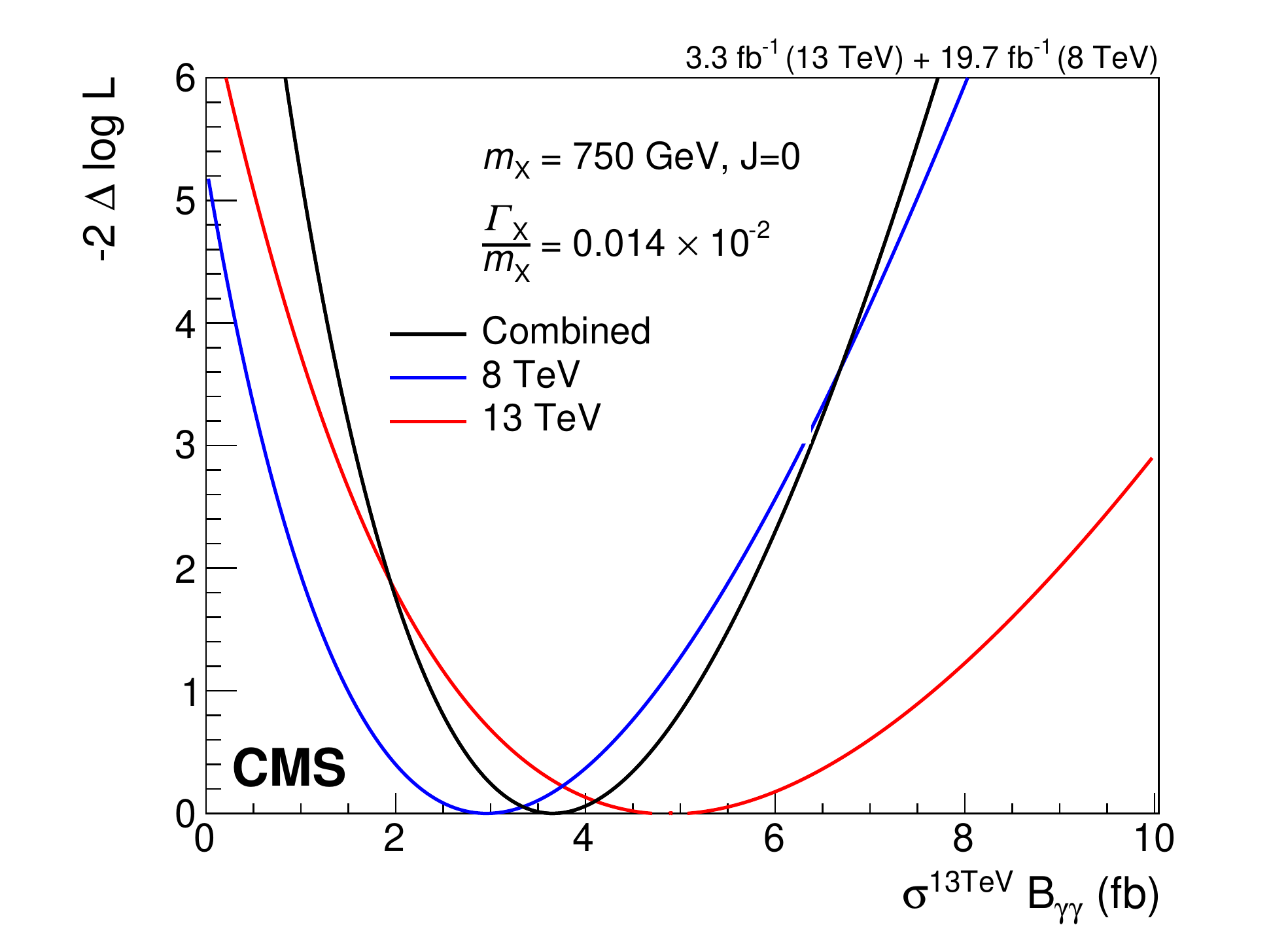}
    \includegraphics[width=\cmsFigWidth]{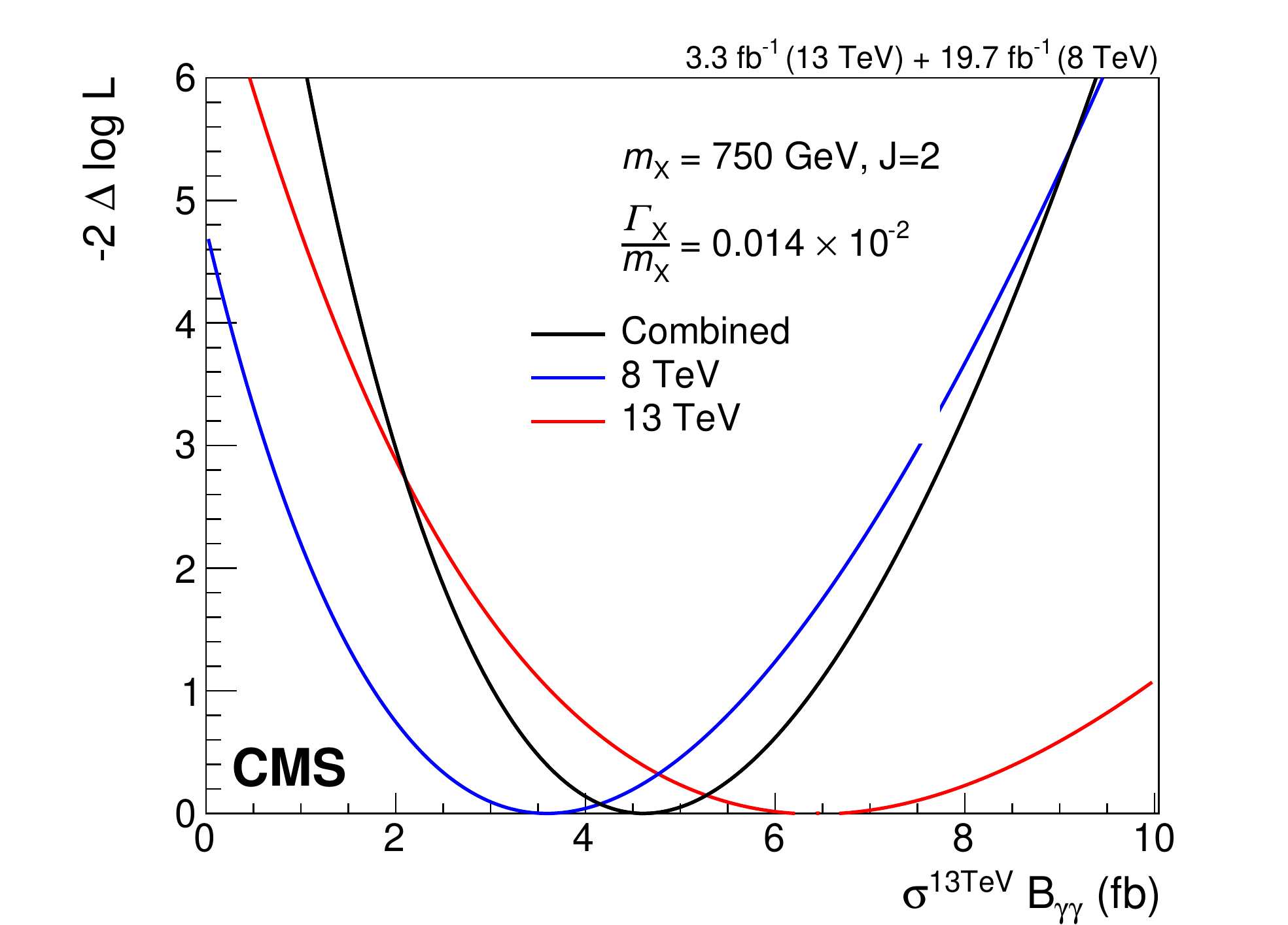}
    \caption{
      Likelihood scan for the cross section corresponding to the largest excess in the
      combined analysis of the 8 and 13\TeV data sets.
      The \cmsLeft (\cmsRight) plot corresponds to the scalar (\RS graviton) signals.
      The 8\TeV results are scaled by the expected ratio of cross sections in each
      scenario.
      \label{fig:extra_comb_mu}
    }
\end{figure}

%% file: EXO-16-018-authorlist.tex
\textbf{Yerevan Physics Institute,  Yerevan,  Armenia}\\*[0pt]
V.~Khachatryan, A.M.~Sirunyan, A.~Tumasyan
\vskip\cmsinstskip
\textbf{Institut f\"{u}r Hochenergiephysik der OeAW,  Wien,  Austria}\\*[0pt]
W.~Adam, E.~Asilar, T.~Bergauer, J.~Brandstetter, E.~Brondolin, M.~Dragicevic, J.~Er\"{o}, M.~Flechl, M.~Friedl, R.~Fr\"{u}hwirth\cmsAuthorMark{1}, V.M.~Ghete, C.~Hartl, N.~H\"{o}rmann, J.~Hrubec, M.~Jeitler\cmsAuthorMark{1}, A.~K\"{o}nig, I.~Kr\"{a}tschmer, D.~Liko, T.~Matsushita, I.~Mikulec, D.~Rabady, N.~Rad, B.~Rahbaran, H.~Rohringer, J.~Schieck\cmsAuthorMark{1}, J.~Strauss, W.~Treberer-Treberspurg, W.~Waltenberger, C.-E.~Wulz\cmsAuthorMark{1}
% ...
\vskip\cmsinstskip
% ...
1:~~Also at Vienna University of Technology, Vienna, Austria\\
% ...
